\documentclass[twocolumn,tighten]{aastex63}

\usepackage{color, hyperref}
\usepackage[normalem]{ulem}
\usepackage{amsmath}
\usepackage{listings}
\usepackage{CJKutf8}
\usepackage[figuresright]{rotating}

\DeclareMathAlphabet\mathbfcal{OMS}{cmsy}{b}{n} 

\newcommand{\deriv}{\ensuremath{{\rm d}}}  

\newcommand{\params}{\ensuremath{\boldsymbol\theta}}
\newcommand{\eparams}{\ensuremath{\boldsymbol\phi}}
\newcommand{\flux}{\ensuremath{\mathbf{F}}}

\newcommand{\errors}{\ensuremath{\boldsymbol\sigma}}
\newcommand{\feh}{\ensuremath{{\rm [Fe/H]}}}
\newcommand{\afe}{\ensuremath{{\rm [\alpha/Fe]}}}

\newcommand{\likelihood}{\ensuremath{\mathcal{L}}}
\newcommand{\prior}{\ensuremath{\pi}}
\newcommand{\posterior}{\ensuremath{\mathcal{P}}}

\newcommand{\T}{\ensuremath{\mathrm{T}}}

\newcommand{\healpix}{\texttt{healpix}}

\usepackage{graphicx}
\DeclareGraphicsExtensions{.png,.pdf}
\graphicspath{{./}{figures/}}

\newcommand{\ms}{\textsc{MINESweeper}}
\newcommand{\brutus}{\textsc{brutus}}
\newcommand{\starhorse}{\textsc{StarHorse}}

\newcommand{\python}{\textsc{Python}}
\newcommand{\astropy}{\textsc{astropy}}
\newcommand{\healpy}{\textsc{healpy}}

\newcommand{\mesa}{\textsc{mesa}}

\newcommand{\scipy}{\textsc{SciPy}}
\newcommand{\numpy}{\textsc{NumPy}}
\newcommand{\matplotlib}{\textsc{matplotlib}}

\newcommand{\corner}{\textsc{corner}}
\newcommand{\gala}{\textsc{Gala}}
\newcommand{\galpy}{\textsc{galpy}}
\newcommand{\crowdsource}{\textsc{crowdsource}}
\newcommand{\allsky}{\textsc{allsky}}

\newcommand{\mist}{\texttt{MIST}}

\newcommand{\ctk}{\texttt{C3K}}
\newcommand{\augustus}{\texttt{Augustus}}
\newcommand{\gold}{\texttt{Augustus-Gold}}
\newcommand{\silver}{\texttt{Augustus-Silver}}

\shorttitle{Mapping the Milky Way in 5-D}
\shortauthors{Speagle et al.}
\begin{document}

\title{Mapping the Milky Way in 5-D with 170 Million Stars}

\author[0000-0003-2573-9832]{Joshua S. Speagle (\begin{CJK*}{UTF8}{bsmi}沈佳士\ignorespacesafterend\end{CJK*})}
\altaffiliation{Banting \& Dunlap Fellow}
\affiliation{Department of Statistical Sciences, University of Toronto, Toronto, ON M5S 3G3, Canada}
\affiliation{David A. Dunlap Department of Astronomy \& Astrophysics, University of Toronto, Toronto, ON M5S 3H4, Canada}
\affiliation{Dunlap Institute for Astronomy \& Astrophysics, University of Toronto, Toronto, ON M5S 3H4, Canada}
\affil{Center for Astrophysics\:\textbar\:Harvard \& Smithsonian, 
60 Garden St., Cambridge, MA 02138, USA}
\email{j.speagle@utoronto.ca}

\author[0000-0002-2250-730X]{Catherine Zucker}
\affil{Center for Astrophysics\:\textbar\:Harvard \& Smithsonian, 
60 Garden St., Cambridge, MA 02138, USA}
\affil{Space Telescope Science Institute, 3700 San Martin Drive, Baltimore, MD 21218, USA}

\author[0000-0002-7846-9787]{Ana Bonaca}
\affil{Center for Astrophysics\:\textbar\:Harvard \& Smithsonian, 
60 Garden St., Cambridge, MA 02138, USA}
\affil{Carnegie Observatories,
813 Santa Barbara Street, Pasadena, California 91101, USA}

\author[0000-0002-1617-8917]{Phillip A. Cargile}
\affil{Center for Astrophysics\:\textbar\:Harvard \& Smithsonian, 
60 Garden St., Cambridge, MA 02138, USA}

\author[0000-0002-9280-7594]{Benjamin D. Johnson}
\affil{Center for Astrophysics\:\textbar\:Harvard \& Smithsonian, 
60 Garden St., Cambridge, MA 02138, USA}

\author[0000-0002-8658-1453]{Angus Beane}
\affil{Center for Astrophysics\:\textbar\:Harvard \& Smithsonian, 
60 Garden St., Cambridge, MA 02138, USA}

\author[0000-0002-1590-8551]{Charlie Conroy}
\affil{Center for Astrophysics\:\textbar\:Harvard \& Smithsonian, 
60 Garden St., Cambridge, MA 02138, USA}

\author[0000-0003-2808-275X]{Douglas P. Finkbeiner}
\affil{Center for Astrophysics\:\textbar\:Harvard \& Smithsonian, 
60 Garden St., Cambridge, MA 02138, USA}
\affil{Department of Physics, 
Harvard University, 17 Oxford St, Cambridge, MA 02138, USA}

\author[0000-0001-5417-2260]{Gregory M. Green}
\affil{Max-Planck-Institut f{\"u}r Astronomie, K{\"o}nigstuhl 17, 
D-69117 Heidelberg, Germany}

\author[0000-0001-5625-5342]{Harshil M. Kamdar}
\affil{Center for Astrophysics\:\textbar\:Harvard \& Smithsonian, 
60 Garden St., Cambridge, MA 02138, USA}

\author[0000-0003-3997-5705]{Rohan Naidu}
\affil{Center for Astrophysics\:\textbar\:Harvard \& Smithsonian, 
60 Garden St., Cambridge, MA 02138, USA}

\author[0000-0003-4996-9069]{Hans-Walter Rix}
\affil{Max-Planck-Institut f{\"u}r Astronomie, K{\"o}nigstuhl 17, 
D-69117 Heidelberg, Germany}

\author[0000-0002-3569-7421]{Edward F. Schlafly}
\affil{Lawrence Livermore National Laboratory, 
7000 East Avenue, Livermore, CA 94550, USA}

\author[0000-0002-4442-5700]{Aaron Dotter}
\affil{Center for Astrophysics\:\textbar\:Harvard \& Smithsonian, 
60 Garden St., Cambridge, MA 02138, USA}

\author[0000-0003-3734-8177]{Gwendolyn Eadie}
\affiliation{Department of Statistical Sciences, University of Toronto, Toronto, ON M5S 3G3, Canada}
\affiliation{David A. Dunlap Department of Astronomy \& Astrophysics, University of Toronto, Toronto, ON M5S 3H4, Canada}

\author[0000-0002-2929-3121]{Daniel J. Eisenstein}
\affil{Center for Astrophysics\:\textbar\:Harvard \& Smithsonian, 
60 Garden St., Cambridge, MA 02138, USA}

\author[0000-0003-1312-0477]{Alyssa A. Goodman}
\affil{Center for Astrophysics\:\textbar\:Harvard \& Smithsonian, 
60 Garden St., Cambridge, MA 02138, USA}
\affil{Radcliffe Institute for Advanced Study, 
Harvard University, 10 Garden St, Cambridge, MA 02138}

\author[0000-0002-6800-5778]{Jiwon Jesse Han}
\affil{Center for Astrophysics\:\textbar\:Harvard \& Smithsonian, 
60 Garden St., Cambridge, MA 02138, USA}


\author[0000-0002-6561-9002]{Andrew K. Saydjari}
\affil{Center for Astrophysics\:\textbar\:Harvard \& Smithsonian, 
60 Garden St., Cambridge, MA 02138, USA}
\affil{Department of Physics, 
Harvard University, 17 Oxford St, Cambridge, MA 02138, USA}

\author[0000-0001-5082-9536]{Yuan-Sen Ting 
(\begin{CJK*}{UTF8}{bsmi}丁源森\ignorespacesafterend\end{CJK*})}
\affil{Institute for Advanced Study, Princeton, NJ 08540, USA}
\affil{Department of Astrophysical Sciences, Princeton University, 
Princeton, NJ 08544, USA}
\affil{Observatories of the Carnegie Institution of Washington, 
813 Santa Barbara Street, Pasadena, CA 91101, USA}
\affil{Research School of Astronomy and Astrophysics, 
Australian National University, Cotter Road, ACT 2611, Canberra, Australia}
\affil{Research School of Computer Science, 
Australian National University, Acton ACT 2601, Australia}

\author[0000-0002-7588-976X]{Ioana A. Zelko}
\affil{Center for Astrophysics\:\textbar\:Harvard \& Smithsonian, 
60 Garden St., Cambridge, MA 02138, USA}

\begin{abstract}
    We present {\augustus}, a catalog of distance, extinction, and stellar parameter 
    estimates to 170 million stars from $14\,{\rm mag} < r < 20\,{\rm mag}$ and with $|b| > 10^\circ$
    drawing on a combination of optical to near-IR photometry from 
    Pan-STARRS, 2MASS, UKIDSS, and unWISE along with
    parallax measurements from \textit{Gaia} DR2 and 3-D dust extinction maps.
    After applying quality cuts, we find 125 million objects have 
    ``high-quality'' posteriors 
    with statistical distance uncertainties of $\lesssim 10\%$
    for objects with well-constrained stellar types. This is a substantial
    improvement over distance estimates derived from \textit{Gaia} parallaxes alone and in
    line with recent results from \citet{anders+19}. We find the fits
    are able to accurately reproduce the de-reddened \textit{Gaia}
    color-magnitude diagram, which serves as a useful consistency check of our
    results. We show that we are able to clearly detect large, kinematically-coherent
    substructures in our data relative to the input priors, 
    including the Monoceros Ring and the Sagittarius stream, 
    attesting to the quality of the catalog.
    Our results are publicly available at 
    \href{https://doi.org/10.7910/DVN/WYMSXV}{doi:10.7910/DVN/WYMSXV}.
    An accompanying interactive visualization can be found at
    \url{http://allsky.s3-website.us-east-2.amazonaws.com}.
\end{abstract}


\section{Introduction} \label{sec:intro}

A central challenge in astronomy is converting
the projected 2-D positions of sources on the sky 
into 3-D maps that we can use to infer properties
about the Universe. This is especially true when studying
the Milky Way, where recent observational advances have opened possibilities
for 3-D mapping across our Galaxy. But many new discoveries 
depend on the fidelity of such 3-D mapping.
Recent work has exploited full phase-space data to
uncover the remnants of a major 
merger $\sim 10\,{\rm Gyr}$ ago 
\citep[e.g.,][]{koppelman+18,belokurov+18,helmi+18,naidu+21}
and a phase-space ``spiral'' \citep[e.g.,][]{antoja+18} while in the
halo accurate phase-space maps of stellar streams 
have begun to constrain the potential of the Galaxy 
\citep[e.g.][]{johnston+99,lawmajewski10,bonacahogg18}.

\begin{figure*}
\begin{center}
\includegraphics[width=\textwidth]{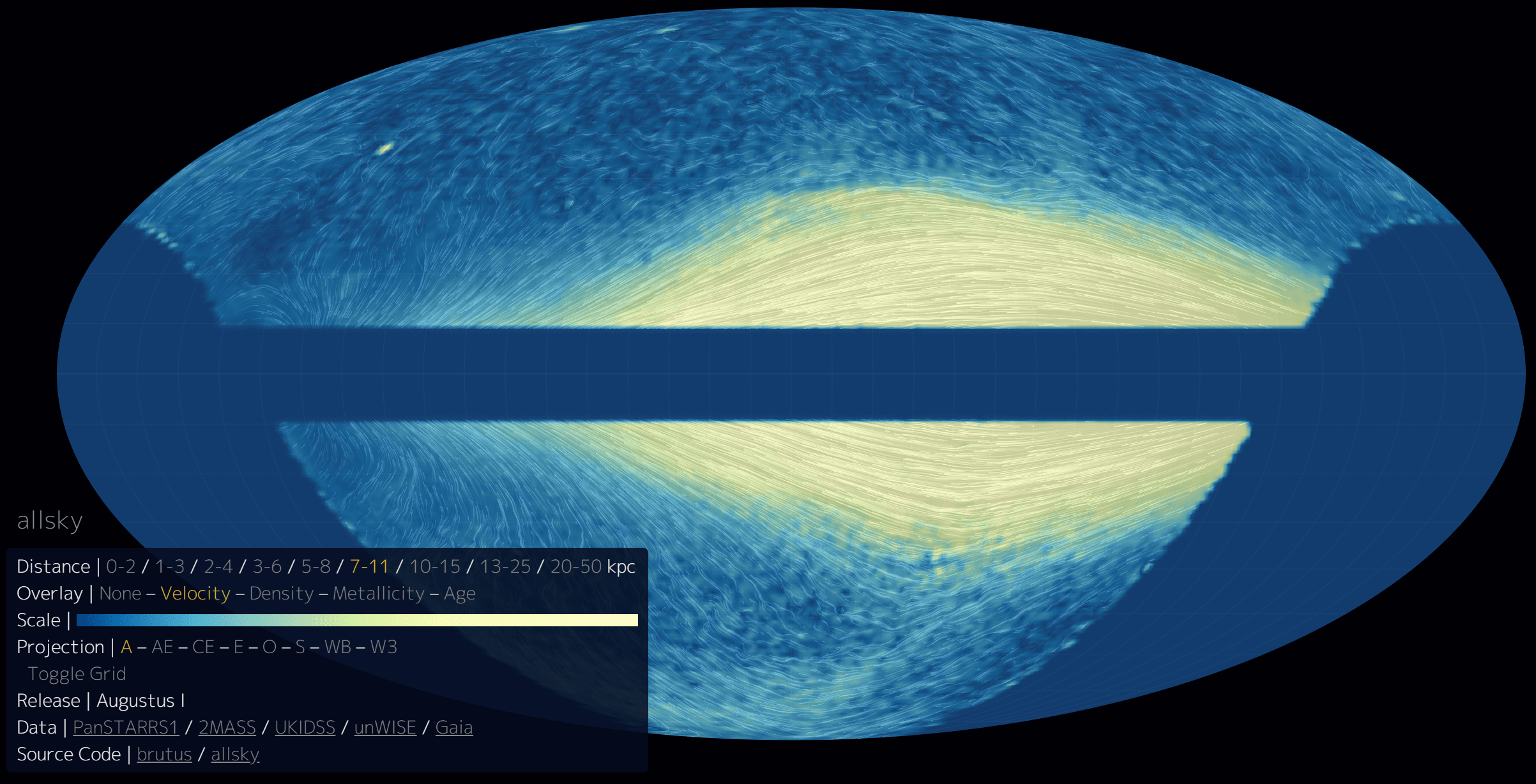}
\end{center}
\caption{A screenshot of an interactive visualization
of 3-D distance and 2-D velocity structure from the {\augustus} catalog
that can be accessed online
at \url{http://allsky.s3-website.us-east-2.amazonaws.com}. The background
color is based on the overall tangential speed of the sources in a given distance bin,
while the white streamlines follow the tangential velocities of the same
sources in the given coordinate projection. 
The interface to change these properties (distance bin, 
background properties, and projection) is shown in the bottom left
and can be opened/minimized by clicking on the ``allsky'' text.
}\label{fig:allsky}
\end{figure*}

These discoveries have benefited from simultaneous
advances across multiple fronts. On the data side,
large missions such as the ground-based
Sloan Digital Sky Survey \citep[SDSS;][]{york+00} and
the space-based \textit{Gaia} mission \citep{gaia+16}
have published enormous public datasets.
Together, these observational efforts promise to 
provide new, much sharper maps of the stellar components of the
Galaxy using billions of individual sources. Simultaneously,
advances in statistical modeling and computational 
power have enabled us to more robustly 
infer the 3-D distribution of a large number of stars 
\citep[e.g.,][]{green+14,bailerjones+18} along with additional
properties such as ages and abundances
\citep[e.g.,][]{ness+15,anders+19,xiang+19,leungbovy19a}. Finally,
advances in numerical simulations and Galactic dynamics
have enabled us to interpret these results in much
more detail \citep[see][and references therein]{rixbovy13,sellwood14,helmi20}.

As most sources ($\sim$ 99\%) seen in large photometric surveys do not have measured
spectra, much of the work associated with deriving
3-D maps to billions of stars has relied on modeling
coarse spectral energy distributions (SEDs)
comprised of flux densities estimated across a range
of broad-band and narrow-band photometric filters 
\citep[e.g.,][]{green+19,anders+19}.
More recently, \textit{Gaia} DR2 \citep{gaia+18}
has also provided astrometric parallax measurements for
many of these sources, giving independent constraints on the distance.

Efforts in this area range from 3-D dust mapping
\citep[e.g.,][]{rezaeikh+18,leikeenslin19,lallement+19,green+19}
to stellar parameter estimation
\citep[e.g.,][]{ness+15,cargile+20,anders+19}.
In Speagle et al. (2021a, subm.), we described new methods implemented
in the public, open-source {\python} package 
{\brutus}\footnote{\url{https://github.com/joshspeagle/brutus}}
that further contribute to these efforts by allowing for quick and robust estimation
of stellar properties, distances, and reddenings to stars with photometric
and astrometric data. In this paper, we present {\augustus}, a ``proof-of-concept''
application of {\brutus} to estimate distances, reddenings, and various stellar properties
to 170 million sources brighter than $r < 20$ mag with Galactic latitudes $|b| > 10^\circ$.

The outline of the paper is as follows.
In \S\ref{sec:data}, we describe the datasets, quality cuts, and selections
used to select the 170 million objects in this work.
In \S\ref{sec:model}, we summarize the approach taken to modeling
and fitting the 170 million sources described with {\brutus}.
In \S\ref{sec:results}, we describe the catalogs produced
by this modeling.
In \S\ref{sec:disc}, we discuss results demonstrating
the quality of the output data, including ``blind''
recovery of the \textit{Gaia} color-magnitude diagram 
and the detection of known large-scale Galactic substructure.
We conclude in \S\ref{sec:conc}. A detailed description
of the data products provided as part of this work
can be found in Appendix \ref{ap:catalog}. An interactive figure
highlighting many of the features of our output catalog can be found
at \url{http://allsky.s3-website.us-east-2.amazonaws.com}; a screenshot is
shown in Figure \ref{fig:allsky} and described in more detail in \S\ref{subsec:allsky}.

\begin{figure*}
\begin{center}
\includegraphics[width=\textwidth]{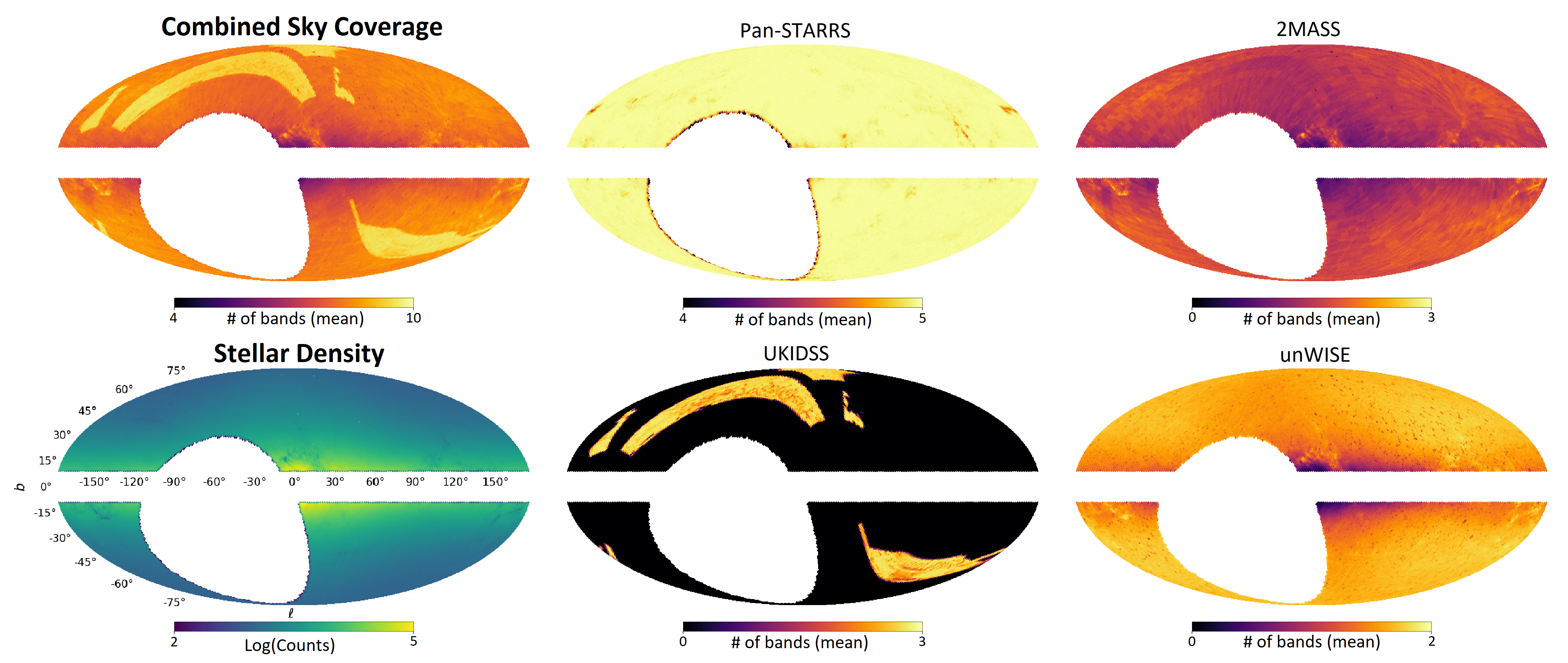}
\end{center}
\caption{Projected maps showing the coverage of the data used in the work
at a {\healpix} resolution of $\texttt{nside} = 64$ plotted as a function
of Galactic longitude $\ell$ and latitude $b$, 
centered on $(\ell, b) = (0^\circ, 0^\circ)$.
The top left panel shows the mean number of bands from all surveys combined,
while the bottom left panel shows the overall (log-)number of stars in each
pixel. The mean number of bands from the individual
surveys are shown in the remaining panels, with Pan-STARRS (\S\ref{subsec:ps})
in the top middle panel, 2MASS (\S\ref{subsec:tmass}) in the top right panel, 
UKIDSS (\S\ref{subsec:ukidss}) in the bottom middle panel, and
unWISE (\S\ref{subsec:unwise}) in the bottom right panel.
While towards the Galactic center we lose a substantial amount
of coverage (transition from orange/yellow to purple in the upper left panel)
due to crowding and dust extinction (transition from blue to yellow in the lower left panel), 
at high Galactic latitudes and in the Galactic outskirts we have $8-10$ bands of
optical to near-IR (NIR) coverage. The uniform coverage of the Pan-STARRS
data is due to the sample selection, which requires $\geq 4$ bands
of Pan-STARRS photometry. Due to the depth of the UKIDSS LAS data,
in regions that overlap with the survey area we have almost
10 bands of wavelength coverage from the optical through the infrared. 
\textit{An interactive version of this figure is available online at
\href{https://faun.rc.fas.harvard.edu/czucker/Paper_Figures/brutus_sky_coverage.html}{this link}.}
}\label{fig:coverage}
\end{figure*}

Throughout the paper, individual parameters (scalars) are notated using standard
italicized math fonts ($\theta$) while vectors and matrices are notated using
boldface ($\boldsymbol{\theta}$). Collections of parameters are notated using sets
($\boldsymbol{\theta} = \{ \theta_i \}_{i=1}^{i=n}$). Vectors should be assumed to
be in column form (i.e. of shape $n \times 1$) unless explicitly stated otherwise.
All magnitudes reported are in the native units 
provided by their corresponding datasets.

\section{Data} \label{sec:data}

Our analysis is based on the combination of several surveys:
\begin{itemize}
    \item the Panoramic Survey Telescope and Rapid Response System 
    \citep[Pan-STARRS;][]{chambers+16},
    \item the Two Micron All Sky Survey \citep[2MASS;][]{skrutskie+06},
    \item the United Kingdom Infrared Telescope Infrared Deep Sky Survey
    \citep[UKIDSS;][]{lawrence+07},
    \item the ``unofficial'' Wide-field Infrared Survey Explorer catalog 
    \citep[unWISE;][]{wright+10,schlafly+19}, and 
    \item the \textit{Gaia} survey \citep{gaia+16}.
\end{itemize}
In \S\ref{subsec:ps}-\S\ref{subsec:gaia}, we describe each of the various datasets.
In \S\ref{subsec:combine}, we describe how the datasets are combined
into a final catalog. The overall sky and wavelength coverage
from these combination of surveys are shown in Figure
\ref{fig:coverage} and \ref{fig:coverage_hist}, respectively.

\subsection{Pan-STARRS} \label{subsec:ps}

The Pan-STARRS survey\footnote{\url{https://panstarrs.stsci.edu/}}
is a multi-epoch, deep, broadband optical
survey of the northern sky visible from 
Haleakala in Hawaii (i.e. $\delta > -30^\circ$).
It observed in five photometric bands ($grizy$) spanning
$0.4\,\micron-1\micron$ with a typical single-epoch $5\sigma$ point-source
exposure depth of $g=22.0\,{\rm mag}$, $r=21.8\,{\rm mag}$, 
$i=21.5\,{\rm mag}$, $z=20.9\,{\rm mag}$, and $y=19.7\,{\rm mag}$
in the AB system \citep{okegunn83}.

The photometry used in this work is based on combined single-epoch photometry
obtained as part of the Pan-STARRS1 $3\pi$ Steradian Survey DR1 \citep{chambers+16}.
Photometry and astrometry were derived from combined images as described
in \citet{magnier+16}. We remove galaxies by requiring that the difference between the
point-spread function (PSF) model photometry and aperture photometry
is $< 0.1\,{\rm mag}$ across at least four bands. We remove objects below
the Pan-STARRS saturation limit in each band, and also objects
with magnitude errors $> 0.2\,{\rm mag}$.

\subsection{2MASS} \label{subsec:tmass}

2MASS\footnote{\url{https://old.ipac.caltech.edu/2mass/}}
is a near-infrared (NIR) survey of the entire sky in three photometric
bands ($JHK_s$) spanning $1\,\micron-2.3\,\micron$ with a typical
$10\sigma$ point-source exposure depth of
$J=15.8\,{\rm mag}$, $H=15.1\,{\rm mag}$, and $K_s=14.3\,{\rm mag}$
in the Vega system.\footnote{While references to the ``Vega system''
suggest a single alternate system, it instead represents a variety of
independent photometric calibrations to differing models of Vega.
This introduces additional systematics when attempting to combine photometry
across different surveys, which will be discussed later.}
We utilize data from the 2MASS “high-reliability” catalog\footnote{
The cuts used in this selection are described at
\url{https://old.ipac.caltech.edu/2mass/releases/allsky/doc/sec2 2.html.}},
which minimizes contamination and confusion by neighboring point and/or
extended sources. We also require errors to be $< 0.2\,{\rm mag}$ and
that no photometry quality (\texttt{ph\_qual}), read quality (\texttt{rd\_qual}),
or galaxy contamination (\texttt{gal\_contam}) flags are set.

\subsection{UKIDSS} \label{subsec:ukidss}

The UKIDSS project\footnote{\url{http://www.ukidss.org/surveys/surveys.html}}
is defined in \citet{lawrence+07}. 
UKIDSS uses the UKIRT Wide Field Camera \citep[WFCAM;][]{casali+07}. 
The photometric system is described in \citet{hewett+06}, and the calibration
is described in \citet{hodgkin+09}. The science archive is described 
in \citet{hambly+08}.

We use data from the UKIDSS Large Area Survey (LAS) second data release
\citep[DR2;]{dye+06,warren+07a,warren+07b}. UKIDSS LAS imaged
4000 square degrees in three fields (an equatorial block, a northern block, and
a southern stripe) that were a subset of the SDSS footprint 
\citep{york+00} in four photometric bands ($YJHK$)
spanning $1\,\micron-2\,\micron$ with a typical $5\sigma$ 
point-source exposure depth of $Y=20.5\,{\rm mag}$,
$J=20.0\,{\rm mag}$, $H=18.8\,{\rm mag}$, and $K=18.4\,{\rm mag}$ in the Vega system.
We require the errors to be $< 0.2\,{\rm mag}$, the
probability of being a star $\texttt{pstar} > 0.9$,
and that no processing/warning errorbit flags
(\texttt{[J\_1/h/k]ppErrBits}) are set.

\subsection{unWISE} \label{subsec:unwise}

The unWISE catalog\footnote{\url{https://catalog.unwise.me/}}
\citep{schlafly+19} is a collection of two billion sources 
observed over the entire sky 
in the infrared (IR) over two bands ($W_1,W_2$) from $3\,\micron-5\,\micron$ 
with the WISE satellite as part of the WISE
and Near Earth Object WISE (NEOWISE) missions.
Compared to the existing AllWISE catalog \citep{cutri+13}, the unWISE
catalog is 0.7 mag deeper due to its use of additional
images from the extended mission along with improved source
modeling in crowded regions using the {\crowdsource} code \citep{schlafly+18}.
This allows it to reach a 50\% completeness point-source depth of
$W_1=17.9\,{\rm mag}$ and $W_2=16.7\,{\rm mag}$ in the Vega system.

We select all objects which are flagged as \texttt{primary} (no duplicate sources),
contain no bitwise quality flags
(\texttt{flags\_unwise}) and that have errors of $< 0.2\,{\rm mag}$.
We also further constrain the fractional flux (\texttt{fracflux})
at an object's given position, which measures the fraction of contamination of light 
from the source due to neighboring objects, to be $> 0.5$
(i.e. a majority of the light at a given position is contributed by
the source being modeled).

\begin{deluxetable}{lcc}
\tablecolumns{3}
\tablecaption{A summary of the photometric offsets
that are \textit{multiplied} to the \textit{observed} flux densities
and the error floors (as a fraction of the flux density) 
that are added in quadrature to the observational uncertainties.
\label{tab:phot}}
\tablehead{Filter & Offset & Error Floor}
\startdata
\cutinhead{Pan-STARRS}
$g$ & 1.01 & 0.02 \\
$r$ & 0.97 & 0.02 \\
$i$ & 0.97 & 0.02 \\
$z$ & 0.96 & 0.02 \\
$y$ & 0.97 & 0.02 \\
\cutinhead{2MASS}
$J$ & 0.99 & 0.03 \\
$H$ & 1.04 & 0.03 \\
$K_s$ & 1.04 & 0.03 \\
\cutinhead{UKIDSS}
$J$ & 0.99 & 0.03 \\
$H$ & 1.04 & 0.03 \\
$K$ & 1.04 & 0.03 \\
\cutinhead{unWISE}
$W1$ & 1.02 & 0.04 \\
$W2$ & 1.03 & 0.04 \\
\enddata
\end{deluxetable}

\begin{deluxetable}{lcc}
\tablecolumns{3}
\tablecaption{Grid of parameters for the {\mist} models
used in this work. See \S\ref{subsec:stars} for additional
details.
\label{tab:mist_grid}}
\tablehead{Minimum & Maximum & Spacing}
\startdata
\cutinhead{\textbf{Initial Mass} ($M_{\rm init}$)}
$0.5\,M_\odot$ & $2.8\,M_\odot$ & $0.02\,M_\odot$ \\
$2.8\,M_\odot$ & $3.0\,M_\odot$ & $0.1\,M_\odot$ \\
$3.0\,M_\odot$ & $8.0\,M_\odot$ & $0.25\,M_\odot$ \\
$8.0\,M_\odot$ & $10.0\,M_\odot$ & $0.5\,M_\odot$ \\
\cutinhead{\textbf{Initial Metallicity} ($\feh_{\rm init}$)}
$-4.0$ & $+0.5$ & $0.06$ \\
\cutinhead{\textbf{Equivalent Evolutionary Point} (${\rm EEP}$)}
$202$ & $454$ & $12$ \\
$454$ & $808$ & $6$ \\
\enddata
\end{deluxetable}

\subsection{\textit{Gaia}} \label{subsec:gaia}

\begin{figure*}
\begin{center}
\includegraphics[width=\textwidth]{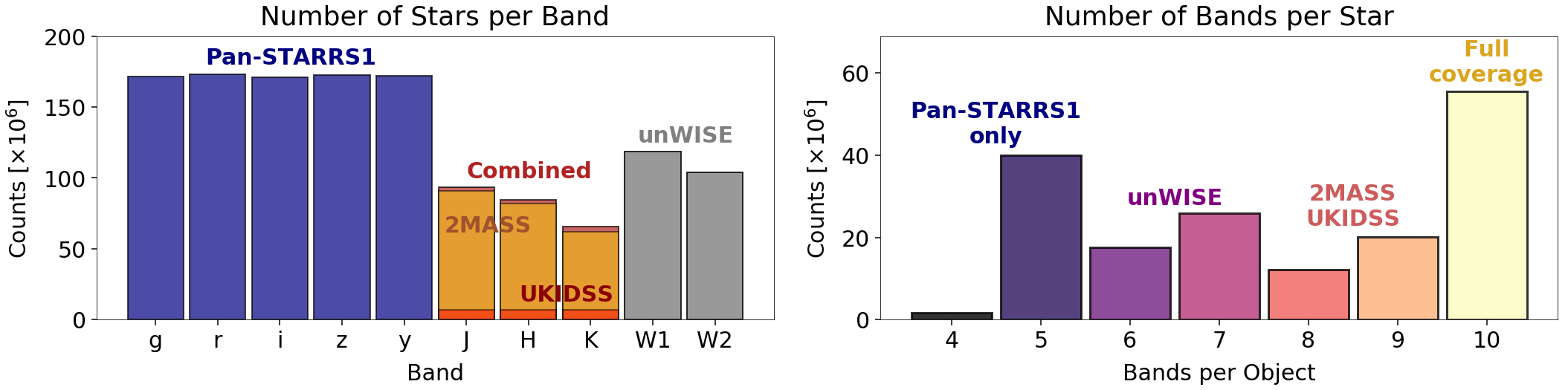}
\end{center}
\caption{Histograms illustrating the wavelength coverage of the data used in
this work that serves to complement Figure \ref{fig:coverage}. 
The left panel shows the number of stars available in each 
band, color coded by survey, illustrating the uniform Pan-STARRS
selection and amount of NIR and IR coverage available through 2MASS
and unWISE, respectively. Compared with Figure \ref{fig:coverage}, 
we can see that although UKIDSS data is quite deep, the
total amount of objects with UKIDSS photometry is quite small
due to its high-latitude targeting. We also see a substantial
amount of objects are detected in the unWISE catalog.
The right panel shows the number of bands per star, highlighting
that the majority of the sample has full coverage across
all 10 possible bands. Regions where we roughly
lose coverage across various surveys are labeled,
with the next largest peak of objects having only
$\sim 5$ bands of coverage (mostly in the optical) closer to the Galactic plane.
\textit{An interactive version of this figure is available online at
\href{https://faun.rc.fas.harvard.edu/czucker/Paper_Figures/brutus_sky_coverage.html}{this link}.}
}\label{fig:coverage_hist}
\end{figure*}

We use data taken from the \textit{Gaia} second data release \citep[DR2;][]{gaia+18},
which provides photometry ($G, BP, RP$) and astrometry (proper motion and parallax)
measurements for over one billion stars, along with radial velocity measurements
for a small fraction of nearby sources. The astrometric catalog
has a 99.875\% completeness point-source depth of $G \approx 21\,{\rm mag}$
in the Vega system \citep{lindegren+18,gaia+18}.
The typical astrometric uncertainty is around $0.7\,{\rm mas}$ for the 
faintest stars and $0.04\,{\rm mas}$ in the bright limit.

In this work, we only incorporate Gaia parallax measurements and their uncertainties
when modeling individual sources. We only use the photometry and proper motions as additional
checks to validate our results, as will be discussed in \S\ref{sec:disc}.
We impose the same quality cuts as recommended in Equation (11) of \citet{lindegren+18},
which requires sources to have:
\begin{itemize}
    \item $G \leq 21$,
    \item $\texttt{visibility\_periods\_used} \geq 6$, and
    \item $\texttt{astrometric\_sigma5d\_max} \leq 1.2\,{\rm mas} \times \gamma(G)$,
\end{itemize}
where $\gamma(G) = \max\left[1, 10^{0.2(G-18)}\right]$.
We do not impose or record astrometric quality information
from any other quantities (e.g., RUWE).

As discussed in \citet{lindegren+18}, there
are numerous systematics present in the parallax measurements
provided as part of \textit{Gaia} DR2. The ones we consider here
are overall zero-point offsets in the parallax measurements
as well as possible underestimates of the provided errors.
Following work by \citet{schonrich+19}, \citet{leungbovy19b},
and \citet{kahn+19}, among others, we add
$0.054\,{\rm mas}$ to all parallaxes and increase the
measurement errors by adding $0.043\,{\rm mas}$ in quadrature with
the reported uncertainties.

\subsection{Assembling the {\augustus} Catalog} \label{subsec:combine}

We cross-matched all sources in Pan-STARRS, 2MASS, UKIDSS, unWISE, and \textit{Gaia} DR2
after applying the cuts described above within a radius of 1 arcsecond, with the closest
source being selected in the presence of multiple matches. This operation was performed
using the Large Survey Database \citep[LSD;][]{juric11}
hosted on the \textit{Cannon} computing cluster at Harvard University and required Pan-STARRS
to be the ``primary'' catalog that all others were matched to. We impose
a minimum error threshold of $0.005\,{\rm mag}$ in all bands after removing any
additional survey-imposed error floors; we also add in our own (larger) error 
floors as described in \S\ref{sec:model}.
In addition, we purposely mask out 2MASS photometry
whenever UKIDSS photometry in the same band is available. This helps
to avoid adding additional noise from the 2MASS observations, which are
substantially shallower than the UKIDSS observations in the same footprint,
and also prevents us ``double-counting'' systematics.\footnote{If we estimate
uncertainties to be dominated by a systematic error $\Delta$, 
observations from $n$ $\sim$\,identical bands with errors floors of $\Delta$ in each
effectively makes the error floor $\Delta/\sqrt{n}$.}

In addition to the cuts described above, we imposed four additional cuts:
\begin{enumerate}
    \item $\geq 4$ bands of photometry in Pan-STARRS,
    \item a parallax measurement in \textit{Gaia} DR2,
    \item $r < 20\,{\rm mag}$ in Pan-STARRS, and
    \item a Galactic latitude of $|b| > 10^\circ$.
\end{enumerate}
The first cut guarantees that we have approximately uniform coverage
in Pan-STARRS and guarantees we have enough photometry ($\geq 4$ bands)
to be able to run {\brutus}, and ensures our spatial coverage matches that of the
3-D dust prior from \citet{green+19}. The second requirement guarantees that all
sources will have parallax measurements (and \textit{Gaia} photometry),
making it straightforward to compare to previous work such as
\citet{anders+19}. The third requirement limits the size of the sample
and helps us remain within the \textit{Gaia} $G \lesssim 21\,{\rm mag}$
completeness limit, making our sample roughly $r$-band magnitude-limited.
Finally, the fourth cut helps to further limit the size of the sample
while avoiding the intense crowding/dust extinction near the Galactic plane.

The final combined catalog, which we refer to as {\augustus},
has roughly \textit{170 million sources}. The distribution
of sources on the sky is shown in Figure \ref{fig:coverage}, while the
wavelength coverage for stars in our sample is shown in Figure \ref{fig:coverage_hist}.
The lack of data in the south is due to the required coverage in Pan-STARRS,
which only allows for $3\pi$ steradians of coverage.
In general, we find a plurality of sources ($\sim 60$ million)
have 10 bands of optical-through-IR photometry across 
Pan-STARRS, 2MASS/UKIDSS, and unWISE, and that only
40 million ($\sim 25\%$) have $\leq 5$ bands of coverage.

\section{Modeling} \label{sec:model}

The modeling approach for constructing this catalog,
{\brutus}, is described in Speagle et al. (2021a, subm.).
We provide a brief summary below.

\begin{figure*}
\begin{center}
\includegraphics[width=\textwidth]{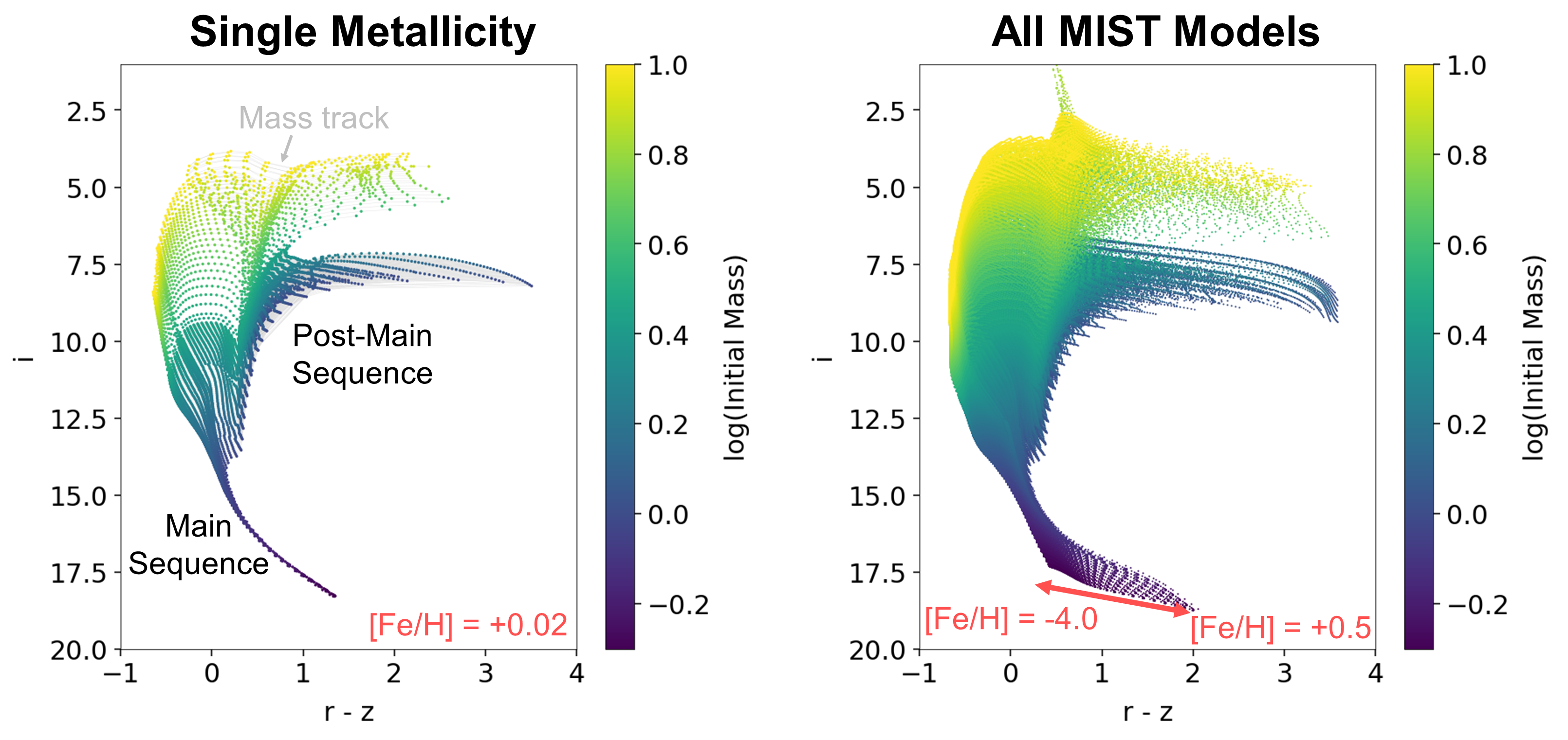}
\end{center}
\caption{Color-magnitude diagrams (CMDs) for the {\mist} models
used in this work as a function of Pan-STARRS $i$-band magnitude
at $d=1\,{\rm kpc}$ and Pan-STARRS $r - z$ color. The left panel
shows the models used with $\feh_{\rm init} = +0.02$,
with the rough locations of Main Sequence
and post-Main Sequence evolutionary phases indicated.
The underlying mass tracks are shown as light grey lines, with
the actual models used shown as points and colored by $\log M_{\rm init}$.
The right panel shows the entire collection of $\sim 750,000$
models defined over $\feh_{\rm init} = -4.0$ to $+0.5$ that are
used in this work.
}\label{fig:mist}
\end{figure*}

\subsection{Statistical Framework} \label{subsec:stats}

{\brutus} uses Bayesian inference to model the
posterior probability $\posterior(\params, \eparams)$
of a set of \textit{intrinsic} stellar parameters
$\params$ (e.g., initial mass $M_{\rm init}$) 
and \textit{extrinsic} stellar parameters (e.g., distance $d$)
as the product of three different components:
\begin{equation}
    \posterior(\params, \eparams) \propto 
    \likelihood_{\rm phot}(\params, \eparams) \,
    \likelihood_{\rm astr}(\eparams) \,
    \prior(\params, \eparams)
\end{equation}

The first term is the \textit{photometric likelihood} between
our model of the flux densities
$\flux(\params, \eparams) \equiv \{ F_i(\params, \eparams) \}_{i=1}^{i=b}$
and the observed flux densities $\hat{\flux} \equiv \{ \hat{F}_i \}_{i=1}^{i=b}$
and their associated errors $\hat{\errors} \equiv \{ \hat{\sigma}_i \}_{i=1}^{i=b}$
across $b$ bands. We assume the data follows
a Normal distribution in each band such that
\begin{equation}
    \likelihood_{\rm phot}(\params, \eparams)
    \equiv \prod_{i=1}^{b} \frac{1}{\sqrt{2\pi\hat{\sigma}_i^2}}
    \exp\left[-\frac{1}{2}\frac{\left(F_i(\params, \eparams) 
    - \hat{F}_i\right)^2}{\hat{\sigma}_i^2}\right]
\end{equation}

The second term is the \textit{astrometric likelihood}, which
compares the predicted parallax $\varpi(\eparams)$ to the
observed value $\hat{\varpi}$ and the associated error
$\hat{\sigma}_\varpi$.\footnote{Note that while other astrometric
measurements such as proper motions are measured by \textit{Gaia}, we do not
model them here.} We also assume the data follow a Normal
distribution such that
\begin{equation}
    \likelihood_{\rm astr}(\eparams) \equiv
    \frac{1}{\sqrt{2\pi\hat{\sigma}_{\varpi}^2}}
    \exp\left[-\frac{1}{2}\frac{\left(\varpi(\eparams) 
    - \hat{\varpi}\right)^2}{\hat{\sigma}_\varpi^2}\right]
\end{equation}

The final term is the \textit{Galactic prior}, which
describes our prior belief over the 3-D distribution
of stars, dust, and their associated properties.
We use the same default prior outlined in Speagle et al. (2021a, subm.).
This includes a thin disk, thick disk, and halo component whose
size/shape are based on \citet{blandhawthorngerhard16} and
\citet{xue+15} with simple, spatially-independent distributions
of initial metallicities $\feh_{\rm init}$ and ages $t_{\rm age}$ as
described in Speagle et al. (2021a, subm.).\footnote{No bulge or bar
component is currently included but will be added in
future work.} The 3-D distribution of dust attenuation $A_V$
is taken to follow the 3-D dust map from \citet{green+19}, 
with variations in the dust curve (as parameterized by $R_V$) taken
to be spatially-independent following that from \citet{schlafly+16}.

\begin{figure*}
\begin{center}
\includegraphics[width=\textwidth]{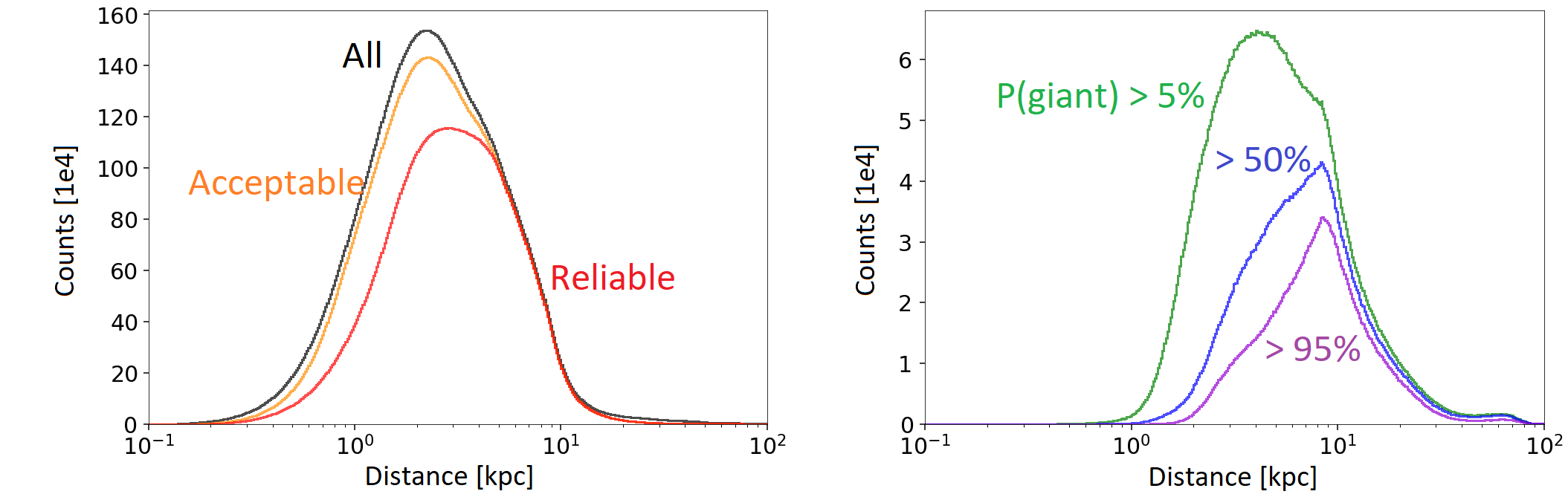}
\end{center}
\caption{A distribution of the distances (taken from a random posterior sample)
for the 170 million objects in our catalog. The left panel shows the number
of objects for the entire sample (black), all the sources with ``acceptable''
fits (the {\silver} subset; orange), and all sources with ``reliable'' posteriors
(the {\gold} subset; red).
There are around 125M sources with reliable posteriors that have distances up
to 10s of kpc. The right panel shows the subset of sources for which the probability
of being a giant (defined as $\log g < 3.5$) is $> 5\%$ (green), $> 50\%$ (blue),
and $>95\%$ (purple). These panels illustrate there are potentially
millions of photometrically-classified giants in the sample, 
although only a few million at very high-confidence.
See \S\ref{sec:results} for more information on the cuts applied here.
}\label{fig:dist}
\end{figure*}

\subsection{Stellar Modeling} \label{subsec:stars}

The stellar models used in this work are the
Modules for Experiments in Stellar Astrophysics
\citep[\mesa;][]{paxton+11,paxton+13,paxton+15,paxton+18,paxton+19}
Isochrone \& Stellar Tracks \citep[\mist;][]{choi+16}. In particular,
we utilize the \textit{non-rotating} models from {\mist} version 1.2.
These are defined in terms of initial mass $M_{\rm init}$,
initial metallicity $\feh_{\rm init}$, and equivalent evolutionary
point \citep[EEP;][]{dotter16}, which correspond to a unique age 
$t_{\rm age}({\rm EEP}|M_{\rm init}, \feh_{\rm init})$ for a given
$M_{\rm init}$ and $\feh_{\rm init}$.

We use three separate pieces to predict the underlying stellar spectrum
$F_\nu(\lambda|\params, \eparams)$.
The first is the {\ctk} stellar atmosphere
models described in \citet{cargile+20} and Speagle et al. (2021a, subm.),
which are computed as a function of effective temperature
$T_{\rm eff}$, log-surface gravity $\log g$, 
surface metallicity $\feh_{\rm surf}$, and surface alpha-abundance
variation $\afe_{\rm surf}$. While version 1.2 of the {\mist} models
provide predictions for $T_{\rm eff}$, $\log g$, and $\feh_{\rm surf}$,
they do not model any variations in $\afe$; as a result, we set
$\afe_{\rm surf} = 0$ by default. The second is a set of
``empirical corrections'' to the {\mist} models based on isochrone 
modeling of nearby open clusters described in Speagle et al. (2021a, subm.).
These are implemented as adjustments to effective temperature $T_{\rm eff}$ 
and stellar radius $\log R_\star$ (and by extension the surface 
gravity $\log g$ and bolometric luminosity
$\log L_{\rm bol}$) as a function of $M_{\rm init}$, with small
modifications as a function of ${\rm EEP}$ and $\feh_{\rm init}$.
The third is a set of dust extinction curves (i.e. reddening laws) from
\citet{fitzpatrick04} to account for the effect of dust extinction.
These are defined as a function of extinction in the $V$ band $A_V$ 
and the ``differential extinction'' $R_V \equiv A_V / (A_B - A_V)$ based on the
ratio of $A_V$ to the difference in extinction $A_B - A_V$
between the $B$ and $V$ bands. Altogether, these give a framework
for generating spectra as a function of
$\params = \{ M_{\rm init}, \feh_{\rm init}, {\rm EEP} \}$
and $\eparams = \{ d, A_V, R_V \}$ with
$\params_* = \{ t_{\rm age}, T_{\rm eff}, \log g, \log R_\star, \log L_{\rm bol} \}$
generated as intermediate values, where $d$ is again the distance to the source.

Finally, to generate the model flux density in a given filter $i$, we integrate the
underlying stellar spectrum $F_\nu(\lambda|\params, \eparams)$
through a filter curve with transmission $T_i(\lambda)$:
\begin{equation}
    F_i(\params, \eparams) = 
    \frac{\int F_\nu(\lambda|\params, \eparams) T_i(\lambda) \lambda^{-1} \deriv\lambda}
    {\int S_\nu(\lambda|\params, \eparams) T_i(\lambda) \lambda^{-1} \deriv\lambda}
\end{equation}
where $S_\nu(\lambda)$ is the source spectrum used to calibrate the
observations. This is the chosen spectrum of Vega in the Vega system and
a constant in the AB system. To avoid having to compute integrals
``on the fly'', we pre-compute photometry over a large grid
of $T_{\rm eff}$, $\feh_{\rm surf}$, $\log g$, $\afe$,
$A_V$, and $R_V$ values in each band, and interpolate over the
resulting photometric predictions in each band using a neural network
as described in Speagle et al. (2021a, subm.). Cross-validation
and hold-out testing suggest that the difference between 
photometry predicted using the neural network versus direct integration
is $\lesssim 0.01\,{\rm mag}$ over a large majority of the parameter space.

\begin{figure*}
\begin{center}
\includegraphics[width=\textwidth]{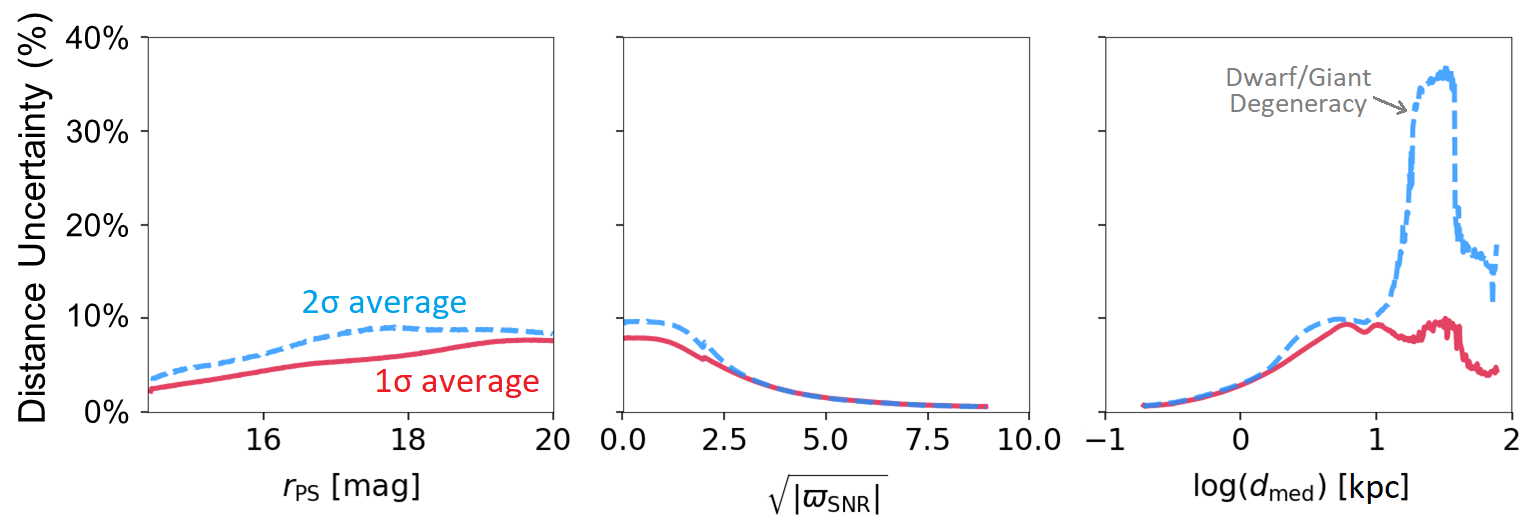}
\end{center}
\caption{Estimates of the \textit{statistical} distance uncertainties
(i.e. excluding systematic uncertainties)
as a function of Pan-STARRS $r$-band magnitude ($r_{\rm PS}$; left), 
parallax signal-to-noise ratio (SNR; middle),
and median estimated distance (right). These are estimated using the distribution
of random distance realizations around the median distance estimates,
with results from half the width of the 68\% credible interval
(``$1\sigma$ average''; solid red) and a quarter the width of the 95\% credible
(``$2\sigma$ average''; dashed blue). We find typical uncertainties of $8-10\%$
at the faintest magnitudes and lowest parallax SNR, with
uncertainties derived from the 95\% CIs
larger than those derived from the 68\% CIs. These differences
can become particularly pronounced at larger distances (right panel) due to
possible degeneracies between nearby dwarf (Main Sequence) 
and faraway giant (post-Main Sequence) stellar evolutionary solutions.
}\label{fig:dist_err}
\end{figure*}

\subsection{Application to Data} \label{subsec:appl}

{\brutus} exploits the nature of the statistical problem to
derive \textit{continuous} estimates of the extrinsic stellar parameters
$\eparams = \{ d, A_V, R_V \}$ over a \textit{grid} of intrinsic
stellar parameters $\params = \{ M_{\rm init}, \feh_{\rm init}, {\rm EEP} \}$.
While interpolating over an input grid of stellar models allows for
smoother probabilistic estimation of underlying parameters \citep{cargile+20}, this
process in general is substantially slower than using pre-computed
grids when the number of parameters being inferred is small $(\lesssim 4)$.
{\brutus} uses grids to exploit this speedup. This leads to ``gridding effects''
that will be visible in subsequent plots shown in this work. 

In brief, {\brutus} fits each object in three steps:
\begin{enumerate}
    \item \textit{Magnitude step}: Compute a ``quick approximation''
    of the solution in magnitudes.
    \item \textit{Flux density step}: Improve the magnitude solution after
    converting the data back to flux density.
    \item \textit{Prior step}: Incorporate information from the
    prior (and the parallax) using Monte Carlo sampling.
\end{enumerate}
These steps are then parallelized across all models in the grid, with
the final posterior estimated using Monte Carlo integration and resampling.
The entire process takes only a few seconds for a typical source with
a mildly-informative parallax measurement.
The grid of stellar models we use in this work (\texttt{grid\_mist\_v8})
are defined in Table \ref{tab:mist_grid} and shown in Figure \ref{fig:mist}
and available online through the {\brutus} GitHub page.

As discussed in Speagle et al. (2021a, subm.) and elsewhere \citep[e.g.,][]{choi+16}, 
there are known systematics offsets between the {\mist} models used in this work
and the photometric data it is being fit to. To account for some of this additional
uncertainty not captured by the empirical corrections described earlier,
we apply zeroth order photometric offsets to the observed flux densities
and increase the effective errors by adding in error floors in quadrature. 
A summary of the offsets and error floors applied in this work can be found in
Table \ref{tab:phot}.

We run {\brutus} using this setup over the 170 million sources in the
{\augustus} catalog using the default hyper-parameters
enabled in \texttt{v0.7.5}\footnote{\url{http://doi.org/10.5281/zenodo.3711493}}.
All computations were performed on the \textit{Cannon} research computing cluster
at Harvard University. Including overheads, the final runtime was
$\sim 700,000$ CPU hours.

\begin{figure*}
\begin{center}
\includegraphics[width=\textwidth]{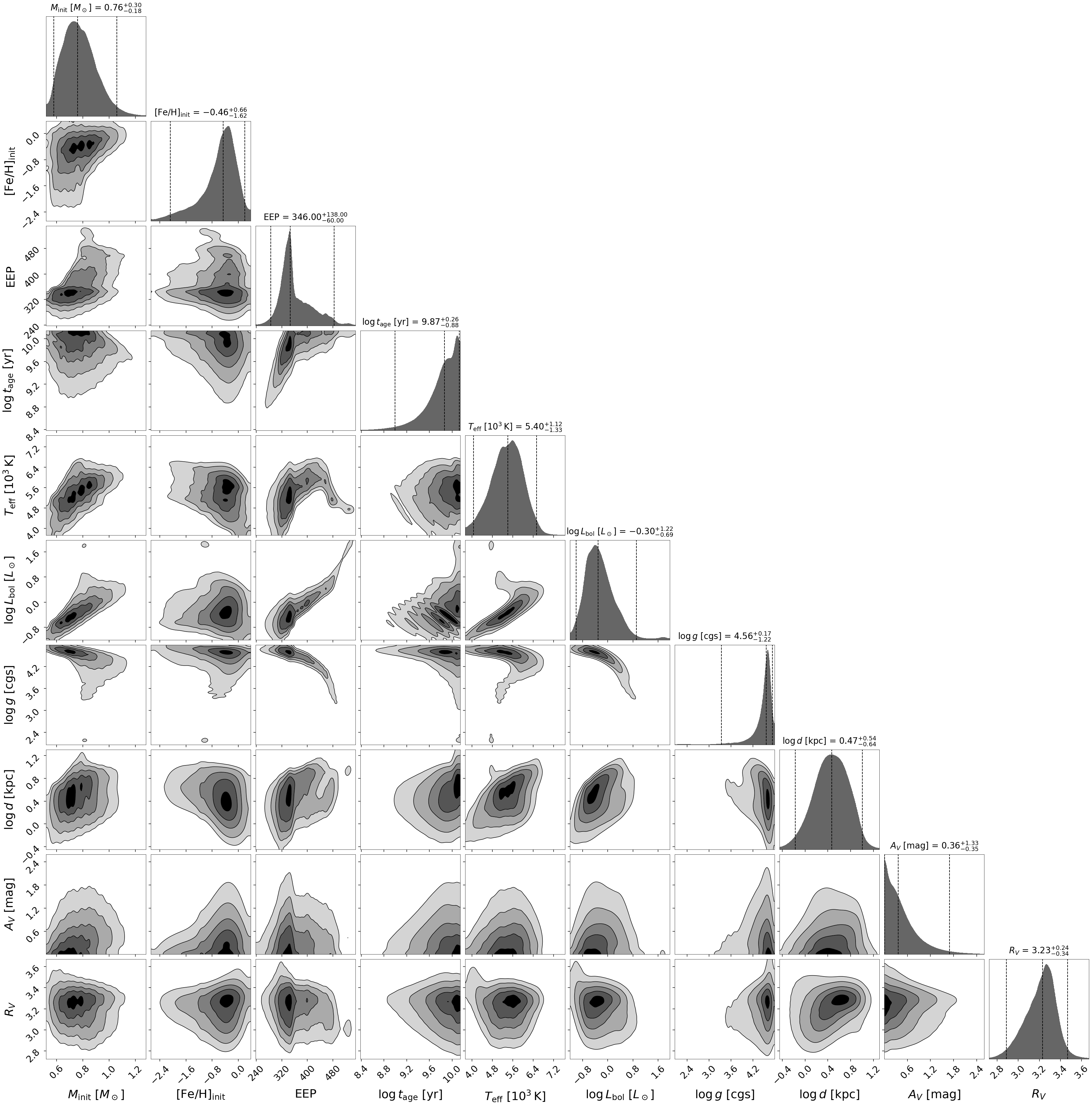}
\end{center}
\caption{A ``corner'' plot showing the collective 1-D and 2-D posterior
distributions for the parameters constrained for
each of the 125 million {\gold} stars. 
These parameters are (from left to right): 
the initial mass ($M_{\rm init}$),
initial metallicity ($\feh_{\rm init}$), equivalent
evolutionary point (${\rm EEP}$), age ($\log t_{\rm age}$),
effective temperature ($\log T_{\rm eff}$), bolometric
luminosity ($\log L_{\rm bol}$), surface gravity
($\log g$), distance $\log d$, extinction ($A_V$),
and ``differential'' extinction ($R_V$). The titles of
each column show the median and the interval encompassing
95\% of the sample. As expected,
the majority the sample comprises sources with 
low extinction ($A_V \lesssim 1$ mag) and initial masses
ranging from $0.55 < M_{\rm init} < 1.2$, with the 
lower limit imposed by the lower $M_{\rm init}$ bound
on our underlying grid of stellar models. 
Gridding effects can be seen in a few panels.
}\label{fig:corner}
\end{figure*}

\section{Catalogs} \label{sec:results}

Using the posterior samples for each object, we post-process the data into two output
catalogs:
\begin{itemize}
    \item a ``point'' catalog\footnote{\href{https://doi.org/10.7910/DVN/WYMSXV}{doi:10.7910/DVN/WYMSXV}}
    containing various information about each object
    and summary statistics describing the stellar parameters, and
    \item a ``samples'' catalog\footnote{\href{https://doi.org/10.7910/DVN/530UYQ}{doi:10.7910/DVN/530UYQ}}
    containing a subset of 25 posterior samples for each object.
\end{itemize}
We will use results from the former catalog 
when highlighting results in this paper; the latter is meant to be used
as a supplement for users interested in additional error modeling.
Detailed descriptions of both catalogs and examples of their usage
can be found in Appendix \ref{ap:catalog}.

Note that after performing most of the computation, we discovered that
\textsc{brutus} \texttt{v0.7.5} contained a bug 
(fixed in more recent versions of the code)
that used the wrong sign when sampling
from correlations between $A_V$ and $R_V$ with distance. We have confirmed
this has a negligible impact on the overall posterior distributions and marginal
distributions for all parameters and therefore 
should not impact the quality of the catalog;
however, it does affect quantities computed directly
from the samples which depend on these quantities (e.g., reddened photometry).

In addition, we found that survey artifacts from
the UKIDSS footprint (deeper near-IR photometry changed the distribution of
stellar parameter estimated) were prominent when projecting results onto the
plane of the sky or Galaxy. As a result, we re-ran all objects in the
UKIDSS LAS footprint without UKIDSS photometry (i.e. using 2MASS instead
when available) and with the same version of the code (\texttt{v0.7.5}) for
consistency; the catalogs for this subset of objects are also provided online
and described in Appendix \ref{ap:catalog}.
While this does degrade the quality of the stellar parameter estimates,
it makes resulting maps more homogeneous. As a result, we opt to use
the ``no UKIDSS'' versions of our results when highlighting results in
this work unless otherwise explicitly stated.

The distribution of a random posterior sample of the distances for each 
star in {\augustus} is shown in Figure \ref{fig:dist}.
We see that the distribution
peaks around a few kpc, with a sharp decline towards larger
distances and a shallower decline towards smaller ones.
The former behavior can be understood by our $r=20$ mag faint
magnitude limit in Pan-STARRS, which makes us primarily sensitive
to giants at larger distances. The latter behavior is due to
a combination of two effects. The first is the increasing differential
volume which goes as $d^2$, which increases 
the raw number of sources available between $d$ and $d + \Delta d$
at larger distances. This counteracts the decreasing
number density of stars as we move away from the Galactic center
and out of the Galactic plane. The second is the $r \sim 14$ mag
saturation limit in Pan-STARRS, which makes us increasingly 
incomplete at nearby distances.

As part of the catalog we generate two quality flags:
\begin{enumerate}
    \item \texttt{FLAG\_FIT}, which diagnoses problems in the
    best-fit model spectral energy distribution (SED), including
    the predicted parallax, and
    \item \texttt{FLAG\_GRID}, which diagnoses when the output
    posterior appears to be artificially truncated by the input model
    parameter grid.
\end{enumerate}
The details of these flags are discussed in Appendix \ref{ap:catalog}.
There are roughly 14 million sources ($\sim 8\%$) which have \texttt{FLAG\_FIT\,=\,TRUE} set,
38 million ($\sim 22\%$) with \texttt{FLAG\_GRID\,=\,TRUE} set, and 47 million ($\sim 27\%$)
with either flag set. We consider the set of roughly
125 million sources with neither flag set to have ``reliable'', high-quality posteriors 
that are sufficient for analysis. We utilize this subset
of stars in {\augustus} in all subsequent analyses and will henceforth refer
to them as the {\gold} subset. Objects with \texttt{FLAG\_FIT\,=\,FALSE} are deemed
``acceptable'' fits, which will henceforth be referred to as the {\silver} subset.

\begin{figure*}
\begin{center}
\includegraphics[width=\textwidth]{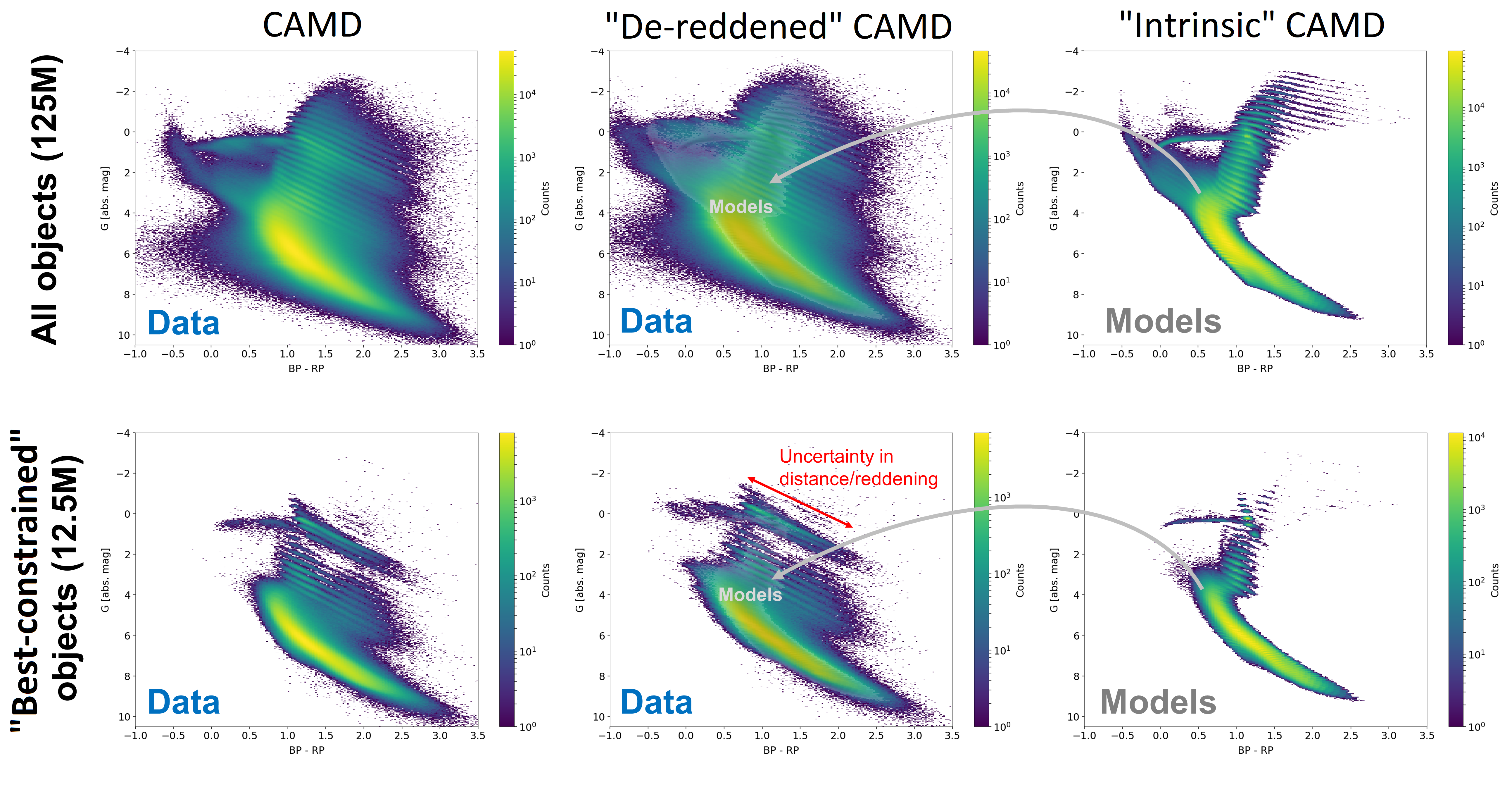}
\end{center}
\caption{The \textit{Gaia} $G$ vs $BP-RP$ color-absolute magnitude diagram (CAMD)
for all sources in {\gold} (125M objects; top) 
and ``best-constrained'' sources (12.5 million objects; bottom)
which have 10 bands of Pan-STARRS, 2MASS, UKIDSS, and/or unWISE
photometry and 95\% distance credible
intervals that are $< 30\%$ of the median distance
(i.e. $|(d_{97.5}-d_{2.5})/d_{50}| < 0.3$). 
The left panels show the CAMD after shifting sources to 
$d = 10\,{\rm pc}$ using a random sample drawn from the stellar posterior.
The middle panels shows the ``de-reddened'' CAMD
using the $A_V$ and $R_V$ values from the same random posterior 
sample and the predicted linear reddening vector
from the stellar parameters associated with them.
The right panels show the predicted CAMD \textit{computed
directly from the models}; this is also over-plotted as the light
gray shaded region in the middle panel for ease of comparison.
As the \textit{Gaia} photometry 
was not used when deriving the stellar posteriors but whose wavelength
coverage overlaps with the Pan-STARRS data, this serves as a useful but limited
check on the internal self-consistency and overall quality of the
results. We find excellent overall agreement between the intrinsic
CAMD predicted by the models and the ``empirical'' CAMD derived from the
data, with uncertainties mostly scattered in the direction of the
reddening vector. Some exceptions to this include
noticeable gridding effects in evolved stellar evolutionary phases 
(thin overdense regions), issues modeling the horizontal giant branch
(upper left regions), and dwarf/giant misclassification (middle right regions).
}\label{fig:gaia_cmd}
\end{figure*}

\begin{figure*}
\begin{center}
\includegraphics[width=\textwidth]{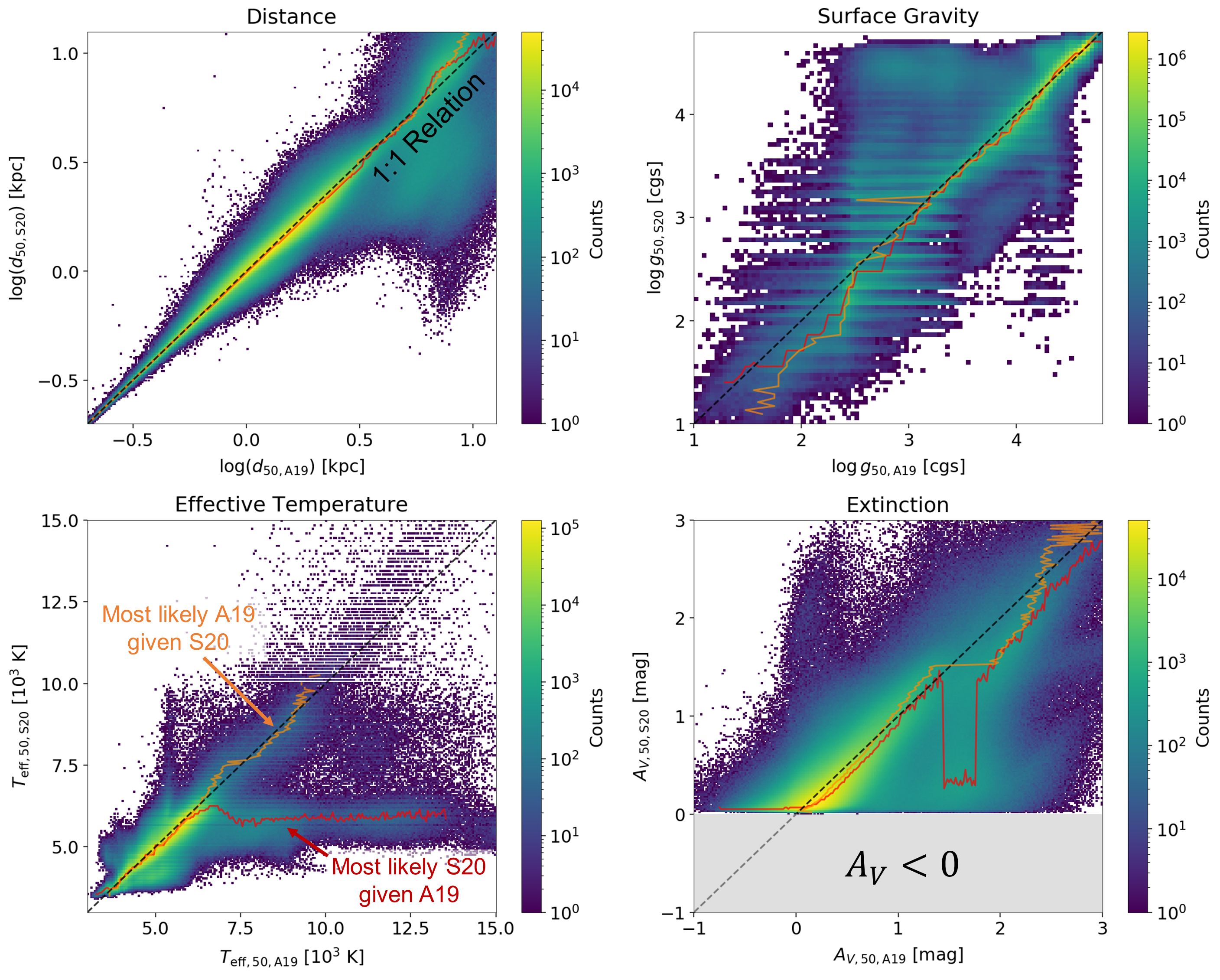}
\end{center}
\caption{Comparisons between median distance (top left),
$\log g$ (top right), $T_{\rm eff}$ (bottom left), and
$A_V$ (bottom right) in this work (\gold; ``S20'') and
\citet{anders+19} (``A19'') for 26 million sources present in both catalogs.
In each panel, the most common (i.e. likely) associated value in {\gold} given
a value from \citet{anders+19} (i.e. $y$ given $x$) is shown as a solid red line, and the
most common associated value from \citet{anders+19} given {\gold} (i.e. $x$ given $y$) is shown as
a solid orange line. A 1:1 relationship is also overplotted as a dashed black curve.
We see that the estimated distances from both catalogs
are extremely consistent with each other across most distances, although
this work generally prefers source below a few kpc to be slightly closer.
The $\log g$ values are also consistent with each other (outside of gridding effects),
although there is a clear excess of sources that are classified as low-mass dwarfs in
{\gold} relative to \citet{anders+19}. The reason for this discrepancy
can be seen when examining the estimated $T_{\rm eff}$, which shows that
while values are consistent across both datasets below $T_{\rm eff} \lesssim 6000\,{\rm K}$,
\citet{anders+19} prefers to make sources substantially hotter than {\gold}
for $T_{\rm eff} \gtrsim 6000\,{\rm K}$.
This leads to a corresponding increase in the number of sources with
$A_V \gtrsim 1$ mag, where the higher reddening combined with the intrinsically
bluer (hotter) colors end up giving similar SEDs as intrinsically 
redder (cooler) sources with less reddening.
Note that the naming convention ``S20'' is based on the fact that the original
analysis was included in the lead author's PhD Thesis, which was
accepted in 2020 and can be found online at 
\url{https://nrs.harvard.edu/URN-3:HUL.INSTREPOS:37365889}.}
\label{fig:starhorse}
\end{figure*}

In Figure \ref{fig:dist}, we show the impact each of these flags
has on the distribution of stellar distances. Overall, we find that most
poor fits tend to happen preferentially at either small distances 
($\lesssim 5\,{\rm kpc}$) or extremely large ones ($\gtrsim 15\,{\rm kpc}$).
Internal investigation reveals this can be due to a variety of failure modes, some
of which are outlined below:
\begin{enumerate}
    \item \textit{Bad photometry}: One outlying band will lead to extremely poor fits.
    This is more common near the Pan-STARRS $r \sim 14\,{\rm mag}$ saturation limit.
    \item \textit{Failed cross-matching}: Multiple objects within the
    same 1 arcsec radius can be inappropriately matched, leading to
    ``mixed'' SEDs that are difficult to model.
    \item \textit{Blending effects}: In crowded regions,
    significant portions of the flux at a given position may be contributed by
    nearby objects, which can impact the measured flux densities 
    for any particular source. This
    becomes stronger at lower $|b|$ values.
    \item \textit{Quasar/galaxy contamination}:
    As discussed in \citet{green+19}, quasars and other point sources
    can often contaminate these samples, especially at fainter magnitudes.
    Since these have very different SEDs compared to our stellar models,
    they often are poorly fit.
    \item \textit{Unresolved binaries}: A non-negligible fraction of sources
    in {\augustus} are expected to be in unresolved binaries 
    \citep[see, e.g.,][]{belokurov+20}. These are not modeled in this work.
    \item \textit{Missing models}: Since our grid only goes down to
    $M_{\rm init} = 0.5\,M_\odot$ and only includes stellar models after they 
    reach the Main Sequence (${\rm EEP} = 202$),
    nearby sources that have $M_{\rm init} \lesssim 0.5\,M_\odot$ or ${\rm EEP} < 202$
    are not part of our grid and therefore will be mismodeled. This can also
    occur if sources fall \textit{between} our grid points, 
    which has coarser spacing on the high-mass end.
    \item \textit{Strong prior disbelief}: 
    As shown in Speagle et al. (2021a, subm.),
    {\brutus} can fail to locate the correct solution if it is sufficiently
    disfavored by the prior. This leads to mismodeling of the SED.
    \item \textit{Imprecise photometric parallaxes}: 
    For nearby sources with extremely high
    signal-to-noise ratio (SNR) astrometric parallax measurements
    from \textit{Gaia} DR2, the model grid in {\brutus} may be too coarse to
    estimate distances with the accuracy needed to match the 
    observed parallax measurements within their measurement uncertainties.
    \item \textit{Heavily extinguished}: The default modeling used in this work
    assumes that $0 \leq A_V \leq 6$. In nearby regions with large $A_V$ (and
    possible large variations in $R_V$), the models will fail to reproduce the
    heavily extinguished SEDs.
\end{enumerate}

By contrast, we find that the vast majority of cases where
\texttt{FLAG\_GRID\,=\,TRUE} occurs when the initial mass $M_{\rm init}$
hits the lower edge of the grid. Imposing this flag therefore imposes a
\textit{de facto} cut on initial mass, limiting the catalog to mostly
sources above $M_{\rm init} \gtrsim 0.55\,M_\odot$. This also explains
why imposing this cut almost exclusively removes sources at smaller
distances, where we are more sensitive to lower-mass objects.

In the right panel of Figure \ref{fig:dist}, we highlight
subsets of {\gold} as a function of the probability that
a source is classified as a ``giant'', which we define as
the probability that it has a $\log g < 3.5$.
The vast majority of stars in our sample ($\sim$ 150 million) are classified
as dwarfs with $P({\rm giant}) \sim 0$, such that even allowing
stars that have only $P({\rm giant}) > 5\%$ only includes roughly
9 million objects. Imposing even stricter criteria such as $P({\rm giant}) > 50\%$
or $P({\rm giant}) > 95\%$ leaves around 5 million and 3.5 million sources, respectively.
Given that {\brutus} is inherently biased against classifying sources
as giants (Speagle et al. 2022a, subm.), we find these to be likely
underestimates of the true number of giants in our catalog. Regardless,
a sample of $> 3$ million giants at high-latitude is already several orders of magnitude
larger than targeted spectroscopic surveys such as the
Hectochelle in the Halo at High Resolution (H3) survey \citep{conroy+19a}.

In Figure \ref{fig:dist_err}, we show the \textit{statistical}
uncertainties in the estimated distances within {\gold} as a function
of Pan-STARRS $r$-band magnitude, parallax SNR $\varpi_{\rm SNR}$,
and median estimated distance. We find typical statistical uncertainties
of $\sim 3-5\%$ near $r \sim 14\,{\rm mag}$ that degrade smoothly
to $8-10\%$ at $r \sim 20\,{\rm mag}$ and $\varpi_{\rm SNR} \lesssim 1$.
Note that these uncertainties \textit{do not} include
systematic uncertainties related to issues with the underlying
stellar models, the assumed Galactic priors, etc.
While in general estimating overall uncertainties from $1\sigma$ and
$2\sigma$ scatter in the distance realizations give consistent
answers, these diverge strongly for median distances
$d_{\rm med} \gtrsim 10\,{\rm kpc}$, at which point
confusion between dwarf and giant solutions lead to multi-modal
distance estimates and largely inflated uncertainties in the tails.
These estimates, especially on the brighter end, are similar to
recent work from \citet{anders+19} and \citet{bailerjones+21} 
and an improvement over the
the purely geometric uncertainties from \citet{bailerjones+18}
and \citet{bailerjones+21} (see \S\ref{subsec:bailerjones}).

The distribution of random posterior samples from
all the stars in {\gold} is shown in Figure \ref{fig:corner}.
As expected for objects with $|b| > 10^\circ$, the vast majority
have $A_V \lesssim 2\,{\rm mag}$ with $>50\%$ of the sample
having $A_V < 0.4$ mag. The metallicity distribution of the
sample peaks around $\feh_{\rm init} \sim -0.5$, similar to that
of the thin disk in our prior, although there is a substantial tail
out to metallicities as low as $\feh_{\rm init} \sim -2.5$.
As expected, most sources have sub-solar initial masses, with
95\% having $0.55\,M_\odot \lesssim M_{\rm init} \lesssim 1.05\,M_\odot$,
and are located on the Main Sequence with ${\rm EEP} < 454$.
We do, however, observe a substantial tail of stars up to and beyond
the Main Sequence turn-off (${\rm EEP} > 454$).

\section{Results and Discussion} \label{sec:disc}

We now wish to highlight some preliminary results illustrating
the quality of the data from the 125 million stars in {\gold}.

\subsection{Reproducing the \textit{Gaia} Color-Magnitude Diagram}
\label{subsec:cmd}

Given known systematics in the theoretical stellar models
used in this and other work \citep[][Speagle et al. 2022a, subm.]{choi+16,anders+19},
one way to examine the reliability of the results is to examine
the ability of the stellar models to reproduce the empirical
color-magnitude diagram (CMD). In order to accomplish this, we purposefully
\textit{did not} use any of the observed \textit{Gaia} DR2 photometry
when computing predictions.
Since these bands are so much broader than the underlying
Pan-STARRS bands that they overlap with, they can serve as a (limited)
\textit{posterior predictive} check on the overall
quality of the fits derived using the Pan-STARRS, 2MASS, UKIDSS, and
unWISE data. In other words, we can test what the
photometry in the \textit{Gaia} bands \textit{should be} (the `predictive')
based on the model constraints imposed from the other fitted bands (the `posterior').

Using the {\brutus} package, we take a random posterior 
sample of the distance $d$, extinction $A_V$, and differential
reddening $R_V$ from each object
and use it to compute the corresponding distance modulus $\mu$
and extinctions $A_G$, $A_{BP}$, $A_{RP}$ in the \textit{Gaia} bands
using the filter curves from \citet{maizapellanizweiler18} as described
in Speagle et al. (2021a, subm.). We then use these to ``de-distance''
and ``de-redden'' the \textit{observed} photometry. We then compare
this ``empirical'' CMD to the ``intrinsic'' CMD predicted directly
from the corresponding model. The results of this exercise
for all sources in {\gold} along with a
subset of 12.5 million sources with full photometric coverage
and reasonably-constrained distances ($< 30\%$ 2$\sigma$ errors)
are shown in Figure \ref{fig:gaia_cmd}. We find excellent 
overall agreement between the intrinsic CMD predicted by the models
and the ``empirical'' CMD derived from the data, with uncertainties 
mostly scattered in the direction of the reddening vector.
Note that this scatter is expected given the incorrect sign of the
covariances between $A_V$ and $R_V$ with distance discussed at the beginning
of \S\ref{sec:results}.
Some exceptions to the excellent overall agreement 
include noticeable gridding effects 
in post-Main Sequence stellar evolutionary phases and higher
masses as well as clear discrepancies properly
modeling the horizontal giant branch.

\subsection{Comparison to {\starhorse}} \label{subsec:starhorse}

We also compare our results to previous work. In particular,
\citet{anders+19}, henceforth ``A19'', use a similar approach to derive distances to
270 million sources with \textit{Gaia} $G < 18\,{\rm mag}$. As in this work, their
approach also involved using a set of theoretical isochrones applied to
similar photometric and astrometric datasets. The main differences
between the two studies are as follows:
\begin{enumerate}
    \item A19 utilizes the PARSEC 1.2 models 
    \citep{bressan+12,chen+14,tang+14}
    rather than the (color-corrected) {\mist} models employed here.
    \item A19 uses the empirical dust curve
    from \citet{schlafly+16} to model $A_V$ variation but no $R_V$ variation.
    Here we use the dust curve from \citet{fitzpatrick04} and
    models variation in both $A_V$ and $R_V$.
    \item A19 uses a different form for the underlying
    Galactic prior that includes an additional bulge component
    and different thin disk, thick disk, and halo properties.
    \item A19 does not apply a 3-D dust prior
    to supply additional constraints on $A_V$.
    \item A19 uses a version of the {\starhorse}
    code \citep{santiago+16,queiroz+18,anders+19} to fit a grid over
    stellar models, $d$, and $A_V$ and evaluates the prior over these data points.
    {\brutus} samples $d$, $A_V$, and $R_V$ values and attempts to integrate
    over the Galactic prior while doing so.
    \item A19 fits photometric data from \textit{Gaia},
    Pan-STARRS, 2MASS, and AllWISE. This work excludes fitting
    photometric \textit{Gaia} data and uses unWISE instead of AllWISE.
    \item A19 applies different \textit{Gaia} DR2
    parallax zero-point corrections and photometric offsets/errors
    compared to this work.
\end{enumerate}

\begin{figure*}
\begin{center}
\includegraphics[width=\textwidth]{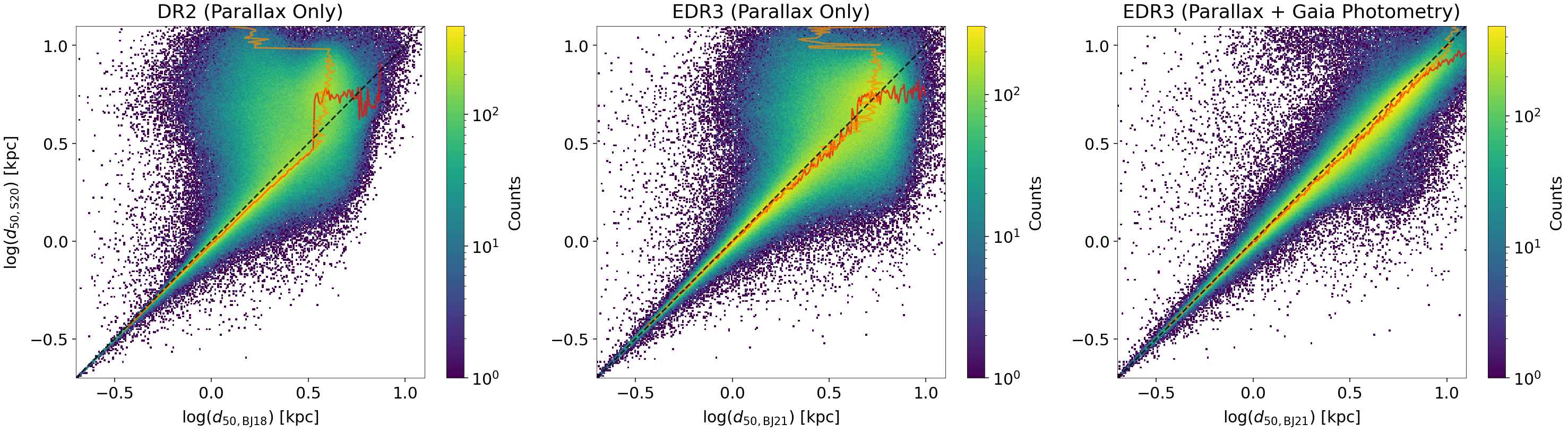}
\end{center}
\caption{As Figure \ref{fig:starhorse}, but now comparing the median distance from this work 
with that derived from \textit{Gaia} DR2 parallaxes only from \citet{bailerjones+18} (BJ18; left)
and from \textit{Gaia} EDR3 parallaxes from \citet{bailerjones+21} (BJ21) without (middle) and with (right)
\textit{Gaia} photometry. The most common (i.e. likely) associated value in {\gold} given
a value from BJ18 or BJ21 (i.e. $y$ given $x$) is shown as a solid red line, and the
most common associated value from BJ18 or BJ21 given {\gold} (i.e. $x$ given $y$) is shown as
a solid orange line. The 1:1 relationship is overplotted as a dashed black curve.
Estimated distances from all three catalogs are in excellent agreement with those derived
here for stars within a few kpc using only the parallax and agree out to further distances
when also considering the BJ21 estimates that also incorporate information from photometry.
Disagreements at larger distances compared to the parallax-only estimates generally arise
due to stronger distance constraints from photometry overwhelming distance estimates for objects with
low parallax signal-to-noise ratios where the Galactic prior tends to dominate the inference.
As with A19, this work prefers sources to be slightly closer than BJ18 and BJ21.}
\label{fig:bailerjones}
\end{figure*}

As A19 performs significant vetting of their
associated catalog across a wide range of surveys,
we want to confirm that we are able to recover similar results 
for sources that overlap between the two catalogs.
After cross-matching sources based on their \textit{Gaia} 
object ID and only selecting ``high-quality'' sources with
\texttt{SH\_GAIAFLAG\_\,=\,000} and \texttt{SH\_OUTFLAG\,=\,00000},
we find roughly 26 million sources that overlap between the
two catalogs. Part of the reason for this small overlap is that
A19 only goes down to $G = 18$ mag rather than the
$r = 20$ mag used in this work; the other main reason is that
this work excludes all sources with $|b| < 10^\circ$ (where the vast
majority of stars actually lie) as well as sources in the 
Southern hemisphere.

\newpage

\onecolumngrid

\begin{sidewaysfigure*}
\begin{center}
\includegraphics[width=\columnwidth]{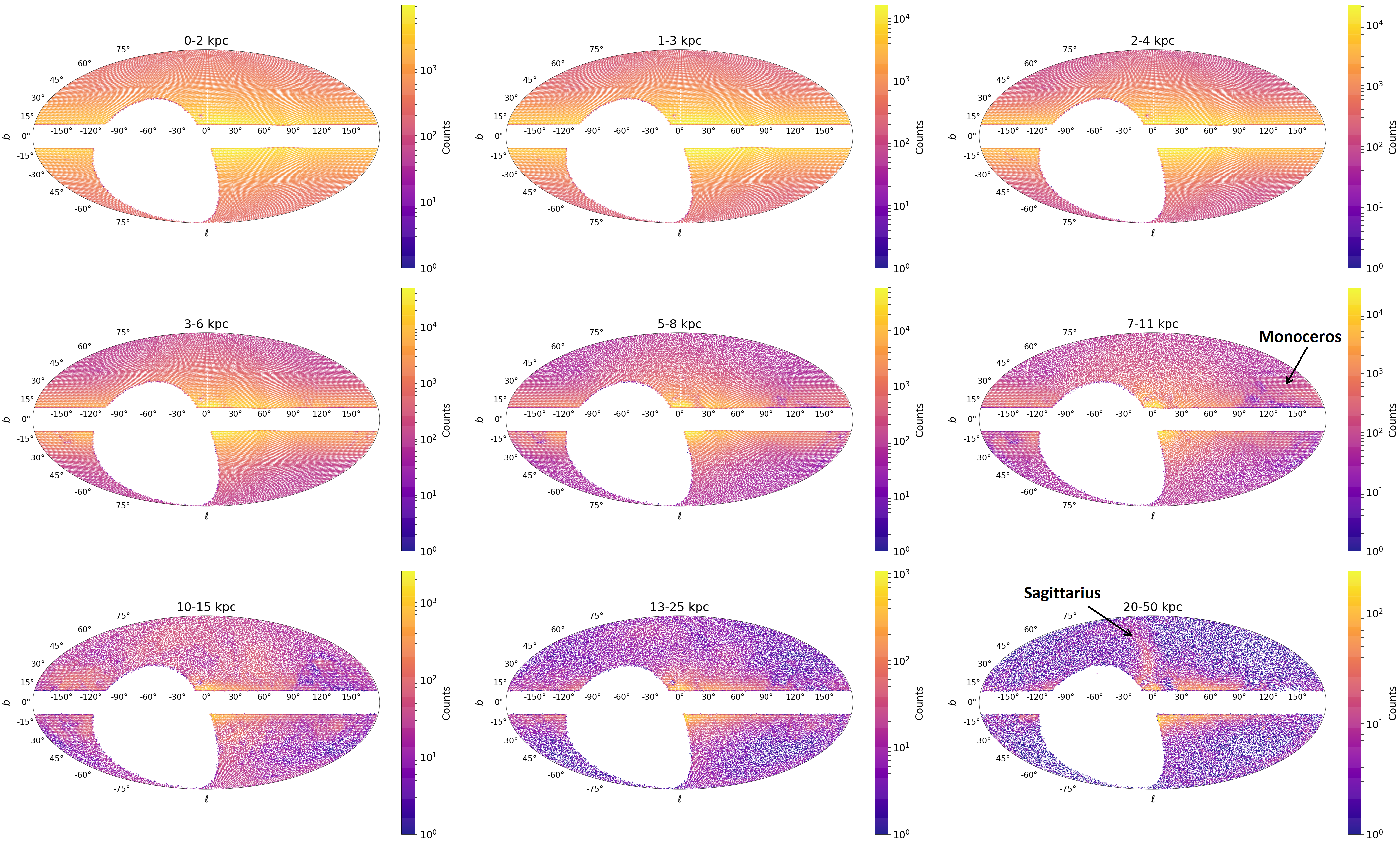}
\end{center}
\caption{The mean number density of stars in (overlapping) distance
bins ranging from $d=0-50\,{\rm kpc}$ at a {\healpix} resolution
of $\texttt{nside} = 64$ (see Figure \ref{fig:coverage}).
The bulk motion of stars within each {\healpix} pixel (corrected
for Solar reflex motion) are indicated
by small arrows and estimated using the mean proper motion.
We can see clear evidence of large-scale features in our
maps, including evidence of the Sagittarius stream (bottom right)
and the Monoceros Ring (middle right) along with ``missing regions''
where substantial foreground dust extinction removes stars from our catalog.
\textit{An interactive version of this figure is available online at
\href{https://faun.rc.fas.harvard.edu/czucker/Paper_Figures/brutus_multipanel_toggle.html}{this link}.}
}\label{fig:quiver_density}
\end{sidewaysfigure*}

\twocolumngrid

\onecolumngrid

\begin{sidewaysfigure*}
\begin{center}
\includegraphics[width=\columnwidth]{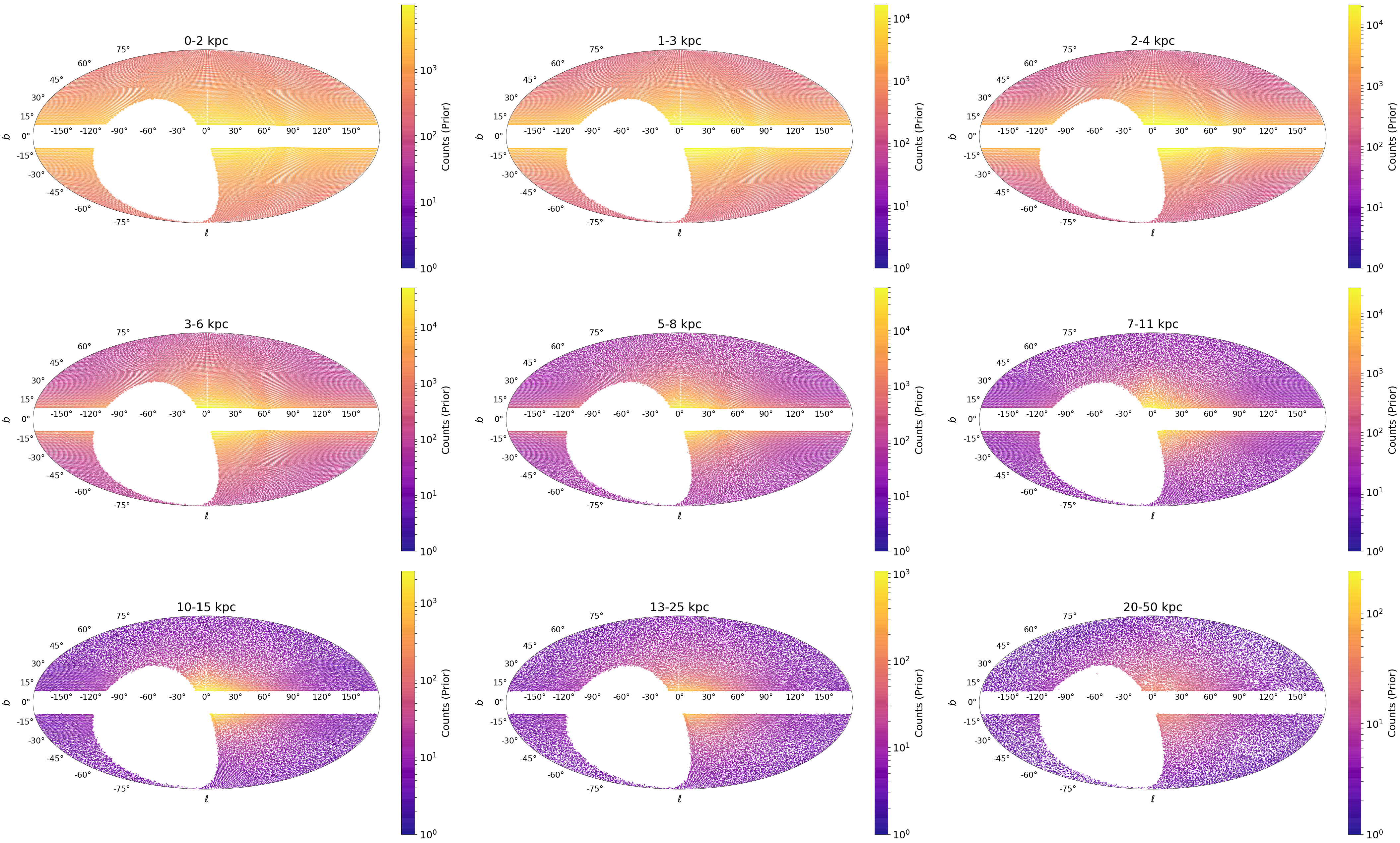}
\end{center}
\caption{As Figure \ref{fig:quiver_density}, but now showing the
expected distribution of sources from the Galactic prior. These
have been normalized so that the total number of sources in each
distance bin are the same in both figures. Since the prior contains only
smooth components, it does not show any evidence for
additional substructure compared to Figure \ref{fig:quiver_density}.
\textit{An interactive version of this figure is available online at
\href{https://faun.rc.fas.harvard.edu/czucker/Paper_Figures/brutus_multipanel_toggle.html}{this link}.}
}\label{fig:quiver_pdensity}
\end{sidewaysfigure*}

\twocolumngrid

\onecolumngrid

\begin{sidewaysfigure*}
\begin{center}
\includegraphics[width=\columnwidth]{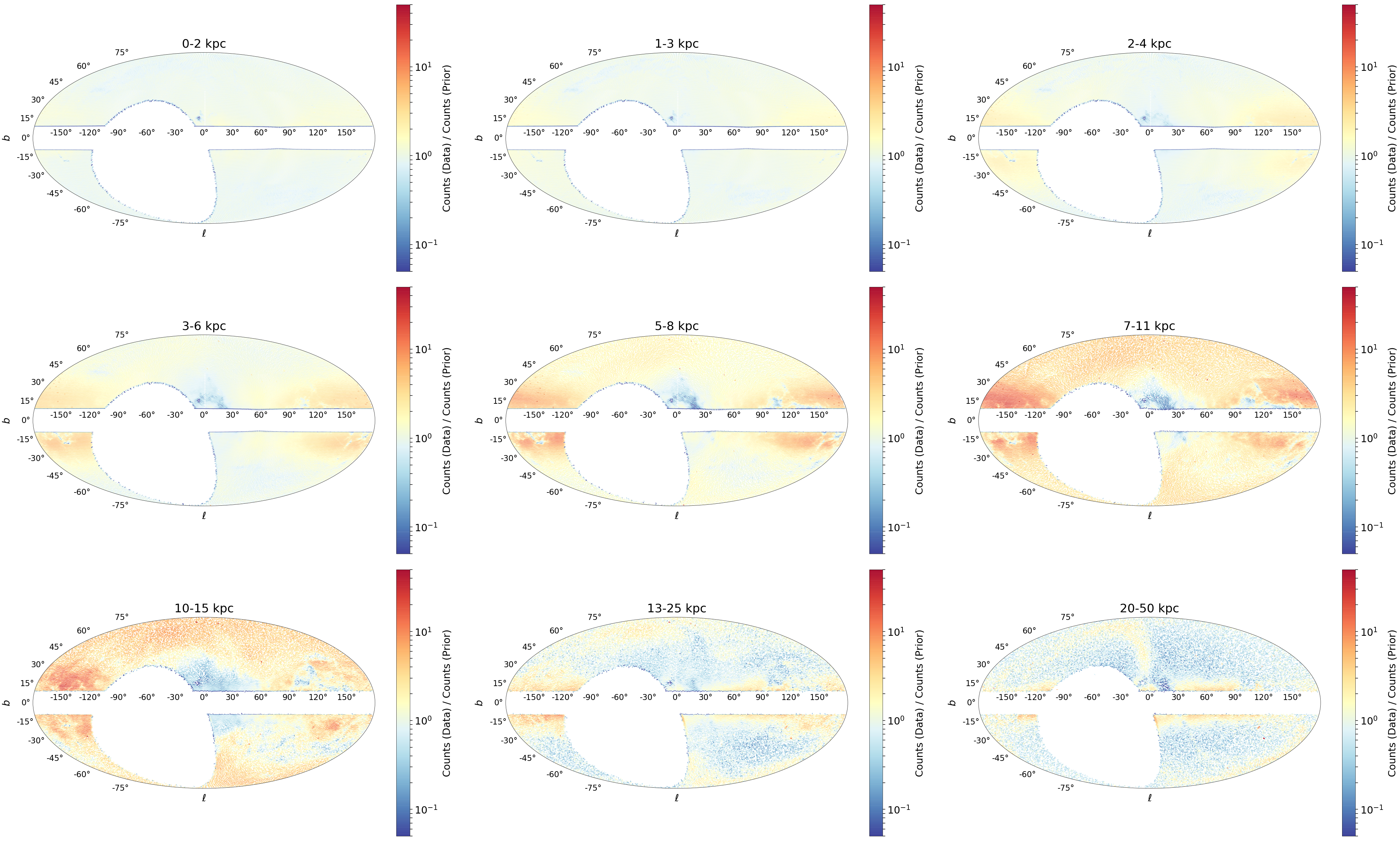}
\end{center}
\caption{As Figure \ref{fig:quiver_density}, but now showing
the ratio between the data and the prior. This highlights
deviations from the assumed Galactic model (i.e. ``the background'')
in order to emphasize substructure present in the data. The Monoceros
Ring in the Galactic anti-center can be clearly seen
as a $\gtrsim 10$ times overdensity relative to the background
from $d \sim 5-15$\,kpc. We also observe broad flaring in the disk, which is not modeled in our prior.
At $d > 20$\,kpc, the Sagittarius stream is also clearly visible, with number densities
also $\gtrsim 10$ times higher than the background.
\textit{An interactive version of this figure is available online at
\href{https://faun.rc.fas.harvard.edu/czucker/Paper_Figures/brutus_multipanel_toggle.html}{this link}.}
}\label{fig:quiver_dpdensity}
\end{sidewaysfigure*}

\twocolumngrid

\onecolumngrid

\begin{sidewaysfigure*}
\begin{center}
\includegraphics[width=\columnwidth]{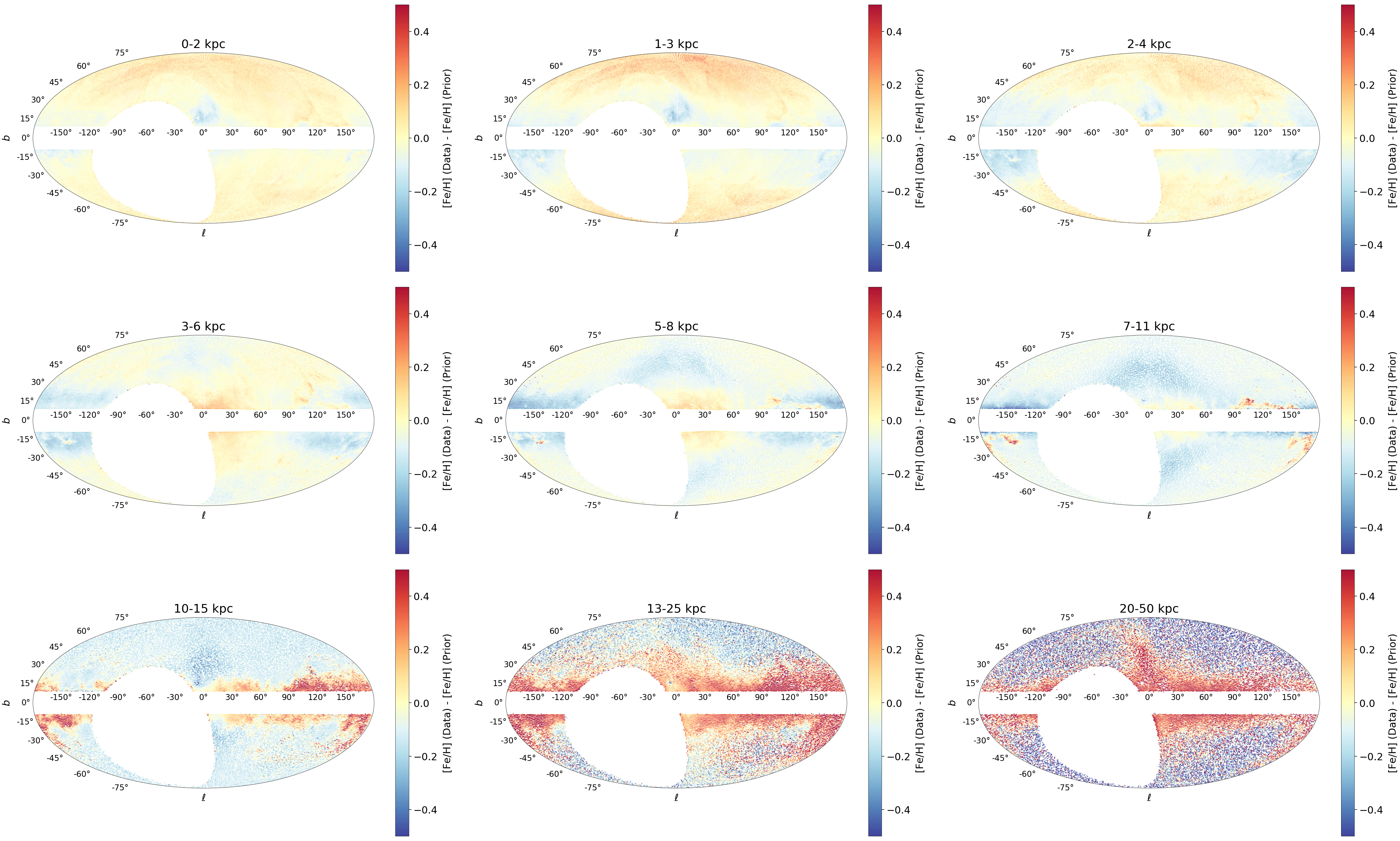}
\end{center}
\caption{As Figure \ref{fig:quiver_dpdensity}, but now showing
the difference between the mean $\feh_{\rm init}$ from the data and the prior.
Deviations can be seen in all distance bins, although the size of the deviations
increases as a function of distance. Clear systematics in $\feh_{\rm init}$ estimation
aligned with foreground dust extinction features and objects near the
Galactic plane are visible. We see evidence for lower-than-expected
$\feh_{\rm init}$ values (compared to the prior) for stars in the Monoceros Ring
and higher-than-expected $\feh_{\rm init}$ values for stars in the Sagittarius stream.
\textit{An interactive version of this figure is available online at
\href{https://faun.rc.fas.harvard.edu/czucker/Paper_Figures/brutus_multipanel_toggle.html}{this link}.}
}\label{fig:quiver_dpfeh}
\end{sidewaysfigure*}

\twocolumngrid

\onecolumngrid

\begin{sidewaysfigure*}
\begin{center}
\includegraphics[width=\columnwidth]{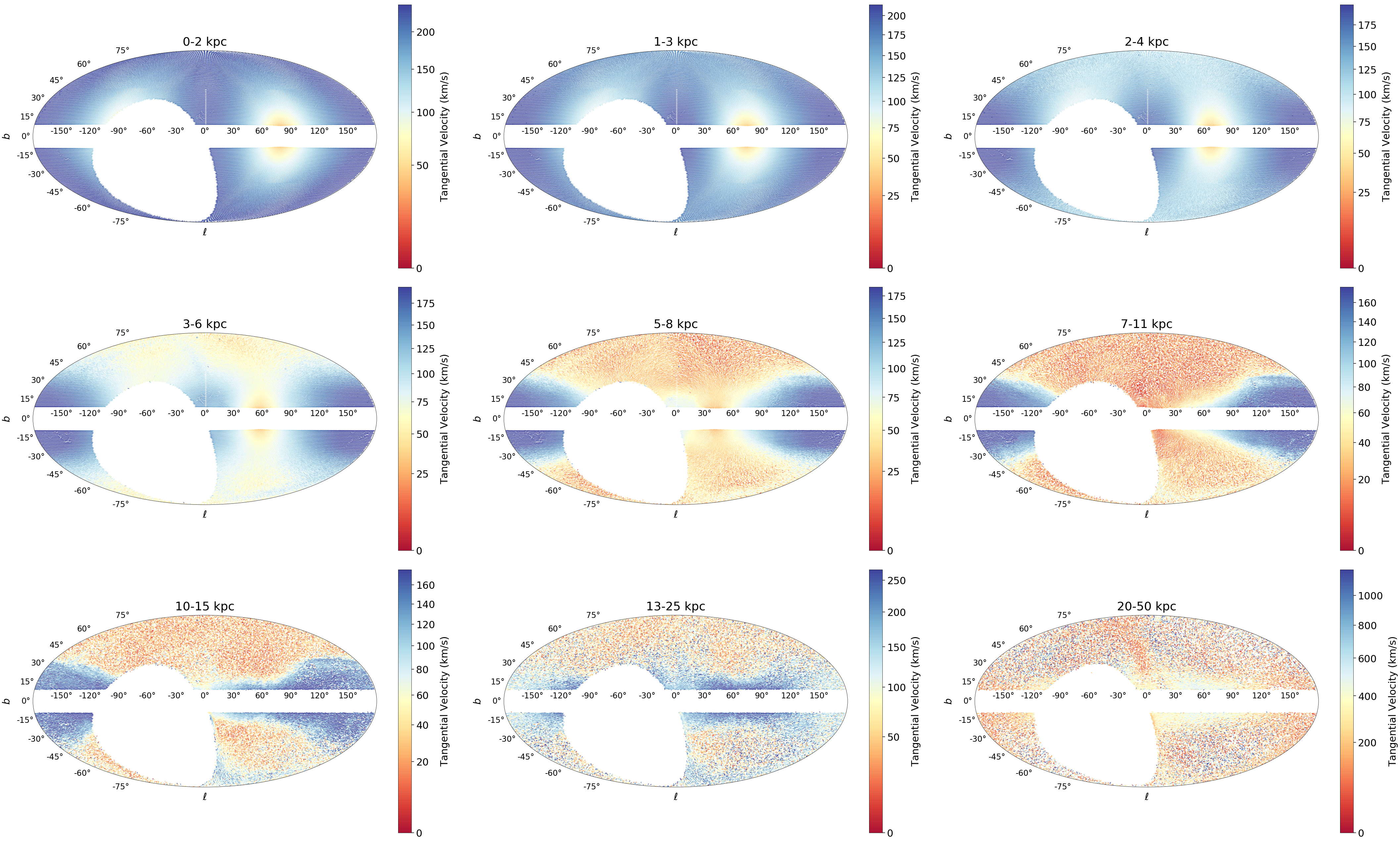}
\end{center}
\caption{As Figure \ref{fig:quiver_density}, but now showing the
mean tangential velocity (in km/s) corrected for
the Solar reflex motion. The overall structure in
the observed velocities at $d \lesssim 6\,{\rm kpc}$ is due
to pointing directly along/opposite the 
expected motion of stars orbiting in the disk.
Kinematically coherent structures associated with known clusters
(which show up as ``outliers'' relative to the background motion of nearby sources)
as well as more extended structures are easily visible.
}\label{fig:quiver_kms}
\end{sidewaysfigure*}

\twocolumngrid

Figure \ref{fig:starhorse} shows the comparison between
the two datasets in a few parameters of interest including
$d$, $\log g$, $T_{\rm eff}$, and $A_V$. Overall, we find
the estimated distances between the two datasets are extremely
consistent with each other, although {\gold} generally
prefers sources nearer than a few kpc to be slightly closer.
This is likely due to small differences in the underlying Galactic prior.
We also find strong agreement between predicted $\log g$ values (outside of
gridding effects due to coarse sampling of post-MS evolutionary phases in this
work), although there is a clear excess of sources that are classified as low-mass dwarfs in
{\gold} relative to A19. 

The reason for this discrepancy
can be seen when examining $T_{\rm eff}$, which shows that
while values are consistent across both datasets below 
$T_{\rm eff} \lesssim 6000\,{\rm K}$, sources in {\gold}
with estimates of $T_{\rm eff} \sim 6000\,{\rm K}$ are much more likely
to have associated estimates of $T_{\rm eff} \gtrsim 8000\,{\rm K}$
in A19. To reproduce the observed SED, these intrinsically
bluer sources need to have more reddening from foreground dust,
leading to an expected increase in higher associated values 
of $A_V$ for some sources in A19 relative to {\gold}.
Note that while {\gold} strictly enforces $A_V \geq 0$ mag and
A19 does not, we find the impact of this choice
does not appear to significantly hamper comparisons other than
around $A_V \sim 0$. We expect that this will slightly bias
distances, intrinsic colors, for stars behind very little dust,
but helps to avoid scenarios that can arise when the
reddening vector is allowed to compensate for systematic color
offsets in the model by exploring non-physical solutions.

\subsection{Comparison to \citet{bailerjones+18,bailerjones+21}} \label{subsec:bailerjones}

In addition to \citet{anders+19}, we also compare our distances to those derived
purely from \textit{Gaia} data based on both DR2 \citep{bailerjones+18} (BJ18) and
EDR3 \citep{bailerjones+21} (BJ21), which was released between constructing the intial catalog
and writing up this manuscript. While BJ18 utilize only parallax information when deriving
their distances, BJ21 include both parallax-only estimates and ones that also
include contributions from an empirical model that incorporates \textit{Gaia} photometry. 
After cross-matching with both catalogs, we find roughly 126 million sources
in common between those samples and the {\gold} sample.

In Figure \ref{fig:bailerjones}, we compare the distance estimates from
BJ18 and BJ21 to those from {\gold}. Overall, we find that excellent agreement between
{\gold}, BJ18, and BJ21 within a few kpc, which further improves when considering estimates
the BJ21 estimates derived including \textit{Gaia} photometry. Given that the BJ18 and BJ21
estimates were derived using substantially different prior assumptions from our model and,
in the case of BJ21, mutually exclusive photometric datasets (as no \textit{Gaia} photometry
was used to derive any stellar properties reported in this work), this agreement 
lends further confidence to the overall accuracy of our distance estimates.

\subsection{Galactic Substructure seen in {\augustus}} \label{subsec:substructure}

As discussed in Speagle et al. (2021a, subm.), inference from photometry
alone is strongly influenced by the underlying Galactic priors and
even including strong constraints from parallaxes can still lead to biases
in inferred stellar properties without tight constraints on $A_V$
(i.e. some knowledge of the intrinsic SED). As such, it is important
to investigate just how much information we are able to recover relative to the prior.

It has been demonstrated in
\citet{green+15}, \citet{chen+19}, \citet{anders+19},
and other work that there is enough information in stellar photometry
to recover distance and extinction estimates to stars with enough precision
to construct detailed, accurate 3-D dust maps. \citet{anders+19}
show that these estimates may also be detailed enough to begin resolving
large-scale features such as the Galactic bar. Given that our map
targets high-latitude regions and goes substantially deeper than that
work, we want to investigate whether we too can (in principle) recover
large-scale substructure purely from astro-photometry alone.

In Figure \ref{fig:quiver_density}, we plot the mean
proper motion directions for all the sources in {\gold} 
with $\texttt{nside} = 64$ {\healpix} resolution, split into
(overlapping) median distance bins ranging from $d_{50}=0\,{\rm kpc}$
to $d_{50}=50\,{\rm kpc}$ and colored by counts in each pixel.
Overall, we see clear evidence of large-scale features in our
maps, including evidence of the Sagittarius stream (bottom right)
and the ``Monoceros Ring'' (middle right) \citep{newberg+02,juric+08,purcell+11,gomez+13,laporte+18a,laporte+18b}.
We also observe issues where systematics clearly play a role in the inferred
stellar properties, especially near the Galactic center and the 
Galactic plane as well as in regions of substantial foreground extinction 
(where we either miss stars entirely or likely somewhat mismodel them).

To get a sense for how significant these features are,
we need to compare them against what we expect given our
Galactic prior. In Figure \ref{fig:quiver_pdensity},
we show the exact same plot except this time
colored by the \textit{expected} number of counts in each pixel,
normalized so that each distance bin contains the same total number of stars.
In this version, we see no evidence for any substructure
in density alone (since our prior includes no kinematic information).
This makes sense, since our prior was, by construction, a \textit{smooth}
model of the Galaxy that did not account for any small-scale structure.

In Figure \ref{fig:quiver_dpdensity} we compare the \textit{ratio}
of the observed number of counts to the expected number of counts.
As expected, we have an under-density of sources in the direction of the Galactic
center, where there is a substantial amount of dust extinction and we are likely
incomplete. We can clearly see the overdensity corresponding to the Monoceros
Ring appears as early as $d \sim 3\,{\rm kpc}$, peaks at $d \sim 9\,{\rm kpc}$,
and extends out to $d \sim 13\,{\rm kpc}$. While there are strong asymmetries
as a function of $\ell$ and $b$, it is difficult to interpret these differences
since these regions have dust clouds along the line of sight.

Beyond $d \sim 15\,{\rm kpc}$, we see clear evidence for the Sagittarius stream,
which has densities at high-latitudes similar to those near the plane.
These densities are in fact so large that they dominate the normalization of the
expected number of counts, leading to these structures being considered 
``normal'' with ratios $\sim 1$ while the rest of the halo is considered ``underdense''.

In Figure \ref{fig:quiver_dpfeh}, we now highlight differences between
the predicted mean $\feh_{\rm init}$ from the Galactic prior 
(which is spatially-independent within each component of our prior) and the mean
$\feh_{{\rm init}, 50}$ derived from the data. Deviations can be seen in all distance bins,
although the size of the deviations increases as a function of distance.
Here we see clear evidence for issues related to survey coverage at nearby
distances with visible survey stripes and at far distance near the Galactic plane. 
There are also clear systematics in $\feh_{\rm init}$ estimation correlated with
foreground dust extinction, with regions with large $A_V$ having
discrepant $\feh_{\rm init}$ relative to background sources. That said,
we do see strong, correlated evidence for lower-than-expected $\feh_{\rm init}$ values 
for stars in the Monoceros Ring and higher-than-expected $\feh_{\rm init}$ values 
for stars in the Sagittarius stream. These findings suggest that we should be able to
use photometric metallicities to explore kinematic and chemical origins of these
structures, such as to follow up on kinematic and chemical separations with other
structures from the Monoceros Ring \citep[e.g.,][]{laporte+20}.

\begin{figure*}
\begin{center}
\includegraphics[width=\textwidth]{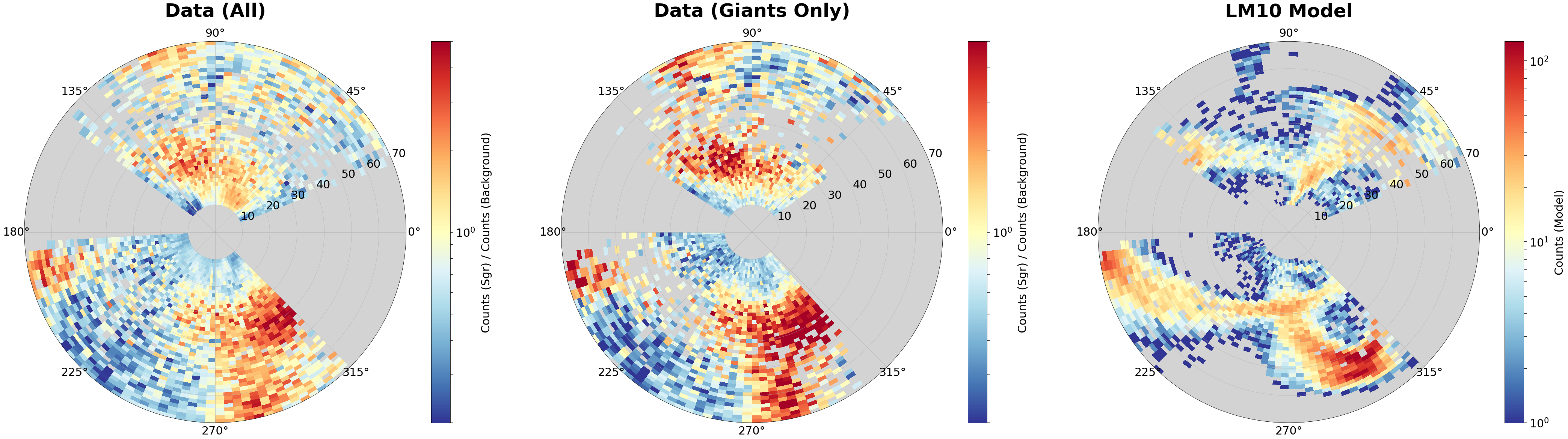}
\end{center}
\caption{The ratio of the observed number density
for objects with $|\beta| < 10$ (``Sagittarius'')
and with $10 < |\beta| < 40$ (``Background'')
as a function of distance and $\Lambda_{\rm Sgr}$, where
$(\Lambda, \beta)$ are coordinates
in the Sagittarius orbital plane. 
The results for all stars with median distances $10\,{\rm kpc} < d_{50} < 70\,{\rm kpc}$
and the subset with $P({\rm giant}) > 95\%$
are shown in the left and middle panels, respectively.
The stellar density from stars taken from the \citet{lawmajewski10}
simulations with $|\beta| < 10$ are shown in the right panel. The broad
correspondence in overall structure between the astro-photometric
distances derived in this work and the results from the simulations
lends confidence that our distances and stellar classifications
are reliable even out to large distances.
}\label{fig:sgr_rad}
\end{figure*}

In Figure \ref{fig:quiver_kms} we plot the associated 
mean tangential velocities in each pixel derived from the median distances
in {\gold} and measured proper motions from \textit{Gaia}, corrected
for the Solar reflex motion using {\astropy} \citep{astropy+13,astropy+18}.
The overall structure in velocities observed at $d \lesssim 6\,{\rm kpc}$
agrees with what we would expect from geometry, where sources orbiting
in the disk are moving directly along/opposite our line of sight.
Kinematically coherent large-scale structure associated with
the Monoceros Ring and Sagittarius stream are clearly visible.
We also clearly see the presence of known large open/globular clusters,
which show up as ``outliers'' in tangential velocity
in a given distance bin relative to the underlying background.
Both of these results are encouraging.

Finally, we examine our detection of the Sagittarius stream
in more detail. We transform 
the coordinates from our sources from Galactic coordinates
$(\ell, b)$ to coordinates aligned with the orbital plane
of the Sagittarius stream $(\Lambda, \beta)$ from \citet{lawmajewski10}.
In Figure \ref{fig:sgr_rad} we try to directly compare these
results with those from simulations taken from \citet{lawmajewski10}
by using the density contrast for sources with $|\beta| < 10$
compared with those with $10 < |\beta| < 30$ as a tracer of the
Sagittarius stream as a function of $d$ and $\Lambda$.
We find our results to be qualitatively consistent regardless
of whether we use all sources or limit ourselves to only those with
$P({\rm giant}) > 95\%$. This broad
correspondence in overall structure between the astro-photometric
distances in {\gold} and the results from the simulations
lends confidence that our distances and stellar classifications
are reliable even out to large distances.

\begin{figure*}
\begin{center}
\includegraphics[width=0.98\textwidth]{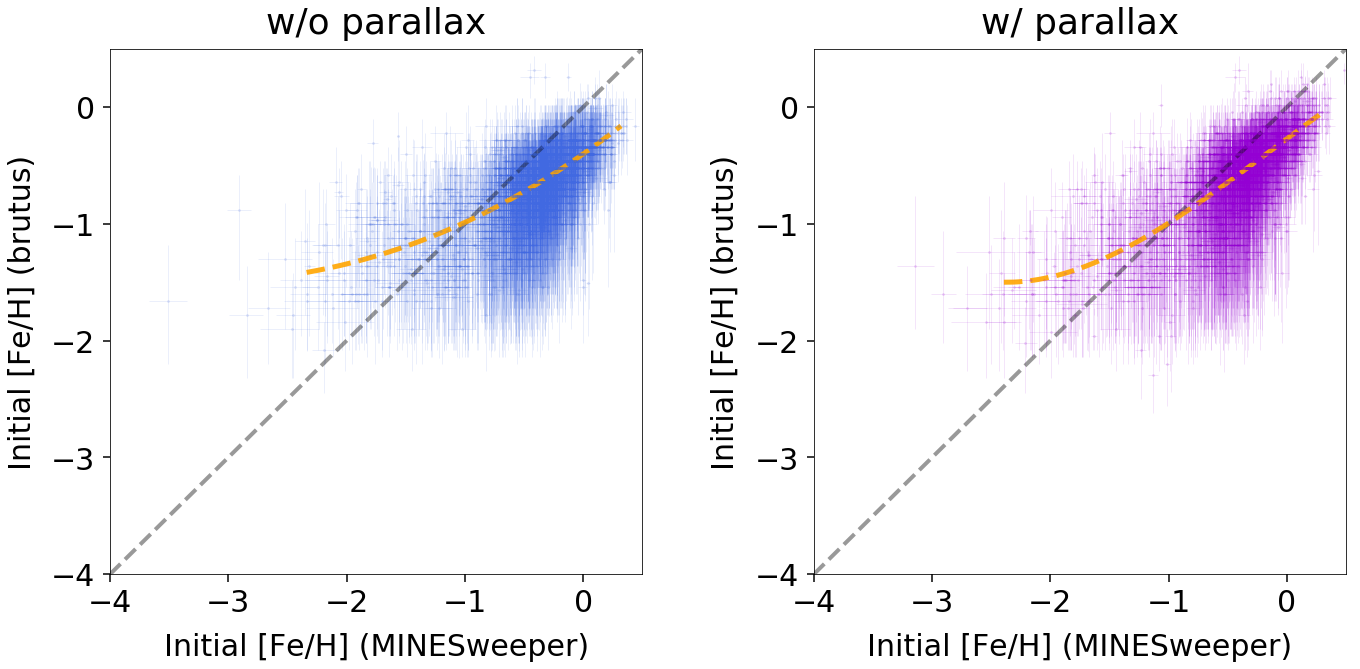}
\end{center}
\caption{A comparison of the metallicities derived from
{\brutus} (used in this work with photometry only) and {\ms} 
(including both photometry and spectroscopy)
for a sample of $n \sim 5100$ objects shown from the H3 Survey
using similar isochrones and photometry without (left, blue) and 
with (right, purple) \textit{Gaia} parallax constraints.
1-sigma errors from both sources are plotted
for each point and the one-to-one relation is shown with a dashed gray line along with
a sliding median (orange dashed line). In both cases, the metallicities derived from
photometry only are found to be substantially biased, although they broadly follow the same trend
as those derived using both spectra and photometry. As discussed in \S\ref{subsec:comments}, this
is due to a fundamental degeneracy where changes in the estimated reddening can be 
accounted for with corresponding changes to the underlying stellar properties, which leads to
increased reliance on the metallicity prior from our Galactic model. These results suggest
our metallicites, along with other derived quantities without good external constraints,
should be used with caution.
}\label{fig:h3_comparison}
\end{figure*}

\subsection{Visualizing 5-D Substructure with {\allsky}} \label{subsec:allsky}

As a final tool to aid exploration and characterization of
the data presented in this work, we modify the public, open
source code \textsc{earth}\footnote{\url{https://earth.nullschool.net/}}
used to visualize wind and ocean currents on the surface of the Earth
to handle the similar types of structure present in 2-D velocities
in a given set of 3-D distance bins. Our public, open source code 
{\allsky}\footnote{\url{https://github.com/joshspeagle/allsky}} is
able to illustrate velocity motion as moving ``streamlines''
following simple non-linear trajectories and allows users
to explore various underlying properties of the data
(velocity, number density, metallicity, etc.) in various projections
(Atlantis, orthographic, equirectangular, etc.) for each
of the nine distance bins shown in this work.
A screenshot illustrating the code is shown in Figure \ref{fig:allsky}.
The full interactive visualization can be accessed online at
\url{http://allsky.s3-website.us-east-2.amazonaws.com} and the data
can be downloaded from the {\allsky} GitHub repository.

\subsection{Additional Remarks} \label{subsec:comments}

While the results described in this section highlight some of the
successes of {\brutus} and the overall quality of the {\gold}
sub-sample, we also want to take some time to explicitly mention
limitations as well as future directions for improvement.
These fall under a few broad categories:
\begin{itemize}
    \item \textit{Limited spatial coverage and depth}: For practical reasons,
    {\augustus} is limited to only covering the Northern sky that
    overlaps with the 3-D dust prior from \citet{green+19}, does
    not include the large amount of objects at $|b| \leq 10^\circ$, and only
    extends down to $r < 20$.
    We aim to break away from these limitations in future work by
    removing our reliance on a previously-estimated 3-D dust prior,
    improving the computational efficiency the underlying {\brutus} code, 
    using larger all-sky datasets such as \textit{Gaia} EDR3 as a base
    to expand off of, and using a broader set of surveys to provide
    deeper optical and (near-)IR photometry in both the North and the South.
    \item \textit{Overly simplistic priors}: The current set of priors
    (see Speagle et al. (2021a, subm.) for additional details) includes
    simple models for a 3-component model that includes a thin disk, thick disk,
    and halo. Not only are all components overly simplistic smooth 
    models (with no substructure, warps, etc.) with spatially-independent 
    metallicity and age constraints, but there is no prior for a bulge or bar.
    We aim to improve on these in future work.
    \item \textit{Low $A_V$ limits}: While the {\augustus}
    catalog targeted the halo, there are still regions where
    extremely high $A_V$ values occur. Stars in these regions
    are inherently mismodeled due to the default maximum value
    of $A_V = 6$ imposed in the catalog. This will be raised to
    much higher values in future work.
    \item \textit{Mass limits:} The theoretical {\mist} isochrones
    we are using have large systematic biases in the predicted photometric colors
    below $M \lesssim 0.5 M_\odot$, 
    even with some of the empirical corrections implemented in
    Speagle et al. (2021a, subm.). This primarily affects our ability to
    model $M$ dwarfs/giants. We hope to use improved isochrone models in future work
    to extend our modeling to lower masses.
    \item \textit{Model-data mismatch}: Offsets between
    the predicted photometry vary systematically across the CMD
    for our given set of isochrones. This imposes systematic limits
    on parameter recovery above the statistical errors present
    in the photometric measurements. We hope the incorporation of
    new data-driven stellar models such as those from
    \citet{green+21} will help to alleviate these issues.
    \item \textit{Gridding effects}: As shown most clearly in
    Figure \ref{fig:starhorse}, the current grid does sample
    a range of surface gravities and effective temperatures but
    still contains visible gridding effects that could impact
    inference. We hope to alleviate this in future work through
    approaches that can apply iterative adaptive refinements.
    \item \textit{Non-Normal errors}: It is possible that the assumption of 
    strictly Normal uncertainties is not valid for the assumed photometric uncertainties
    in both magnitude and flux density, leading to incorrect (likely underestimated) 
    statistical uncertainties in derived properties.
    While systematic effects from survey photometric pipelines
    may lead to empirical error distributions with broader tails (e.g., such as the 
    complex photometric processing for data in \textit{Gaia} DR2; \citealt{gaia+18}), in general
    most reported uncertainties in derived properties are dominated by the effects of 
    model-data mismatch. Similarly, while distribution of statistical parallax uncertainties
    for \textit{Gaia} DR2 appear to follow a Normal distribution out to several standard
    deviations, the applied zero-point corrections can lead to
    systematic offsets that are distinctly non-Normal in nature \citep{lindegren+18}.
    \item \textit{No kinematic constraints}: As can be seen in some
    panels in Figure \ref{fig:quiver_kms}, the predicted tangential velocities
    for some sources indicate clear mismodeling of the distance.
    Especially given the recent improved parallaxes and proper motions from
    \textit{Gaia} EDR3, in future work we hope to incorporate additional kinematic
    constraints to better resolve some of these degeneracies.
\end{itemize}

Although the primary purpose of this catalog is to provide
distance and reddening estimates, we also want to explicitly highlight
possible deficiencies in secondary derived quantities, namely photometric metallicities. 
In Figure \ref{fig:h3_comparison}, we show the metallicity recovery for a subset of
$n \sim 5100$ stars from the H3 survey \citep{conroy+19a}.
In brief, the H3 Survey is a high-latitude
($|b| > 30^\circ$), high-resolution ($R=32,000$) spectroscopic
survey of the distant ($d \gtrsim 2\,{\rm kpc}$ Galaxy. 
Targets are selected purely on their Gaia parallax 
($\varpi < 0.4-0.5\,{\rm mas}$), brightness ($15 < r_{\rm PS1} < 18$), 
and accessibility to the 6.5\,m MMT in Arizona, USA (${\rm dec} > -20^\circ$). 
The survey measures radial velocities to $0.1$\,km/s precision, surface abundances
($\feh_{\rm surf}$ and $\afe_{\rm surf}$) to $0.1$\,dex precision, and
spectrophotometric distances to $10\%$ precision using {\ms} \citep{cargile+20}.
In addition to being a representative, low-reddening subsample of sources,
all derived quantities were estimated using the same
underlying {\mist} isochrones (excluding the empirical corrections
and photometric offsets derived in this work). This makes the estimated metallicities
comparisons both independent (derived using different
codebases and with/without spectra) while still remaining
internally consistent (using similar photometry and underlying stellar models).

As seen in Figure \ref{fig:h3_comparison}, the parameter recovery is substantially
biased and prior-dominated, even with reasonable signal-to-noise photometry and
parallax measurements. As discussed in Speagle et al. (2022a, subm.), 
this is because there is an intrinsic degeneracy in reddened stellar colors 
that only can be broken if the dust attenuation is known to high precision.
Without this, it is possible to shift the brightness and colors of stars by adjusting
dust attenuation (which affects the observed star colors) along with metallicity and 
other stellar properties (which affects the intrinsic star colors). As a result,
estimates are generally dragged towards our prior means for both the thin/thick disk
and halo populations, which shows up as a bias for both high and low-metallicity
objects. Since distance estimates don't break these color degeneracies, this effect
is present when modelling objects with or without parallax constraints, emphasizing the
prior is actually dominating much of the inference.
While internal tests and comparisons presented here imply this bias
doesn't substantially impact our distance estimates, it does mean that our metallicity
estimates, and other prior-dominated derived quantities, 
should be used with caution in any downstream analyses.

\section{Conclusion} \label{sec:conc}

As large-area surveys such as SDSS \citep{york+00},
Pan-STARRS \citep{chambers+16}, 
\textit{Gaia} \citep{gaia+16}, and the Legacy Survey of
Space and Time \citep[LSST;][]{ivezic+08} continue to or
promise to soon provide provide measurements to billions of stars,
the challenge of transforming observations of the projected 
2-D positions of sources on the sky into full 3-D maps
becomes ever more pressing when trying to study the
Milky Way. In this work, we presented results applying
{\brutus} (Speagle et al. 2022a, subm.) --
a public, open-source {\python}
package that uses a combination of statistical
approaches to infer stellar properties, distances,
and extinctions for sources using photometry and astrometry --
to a catalog of 170 million sources ({\augustus}) at 
high Galactic latitudes ($|b| > 10$ mag) down to
$r < 20$ mag with data from Pan-STARRS, \textit{Gaia}, 2MASS, 
UKIDSS, and unWISE.

We find 125 million objects ({\gold}) have good fits and reliable
posteriors with estimated \textit{statistical}
distance uncertainties of $\sim 3-5\%$
at $r=14$ mag to $\sim 8-10\%$ at $r=20$ mag. We show that
our results are able to \textit{predict} the ``empirical'',
de-reddened \textit{Gaia} CMD based on astro-photometric modeling
in other bands, and that the derived stellar parameters
are in excellent agreement with similar results derived in
\citet{anders+19}. We then illustrate the quality
of the data by highlighting its ability to recover
large and small-scale Galactic substructure such as
the Monoceros Ring at $d \sim 10\,{\rm kpc}$ and the
Sagittarius Stream at $d \sim 25\,{\rm kpc}$ in density,
metallicity, and kinematics relative to expectations from the
underlying Galactic prior. Finally, we present an
interactive visualization ({\allsky}) that is able to
highlighted limited 5-D distance and tangential velocity
structure present in our data.

Catalogs summarizing our results are publicly available at the 
Harvard Dataverse\footnote{
\url{https://dataverse.harvard.edu/dataverse/brutus_augustus}} 
and summarized in Appendix \ref{ap:catalog}.
Overall, we hope that our results serve as a useful
value-added catalog that highlight the power
of combined astro-photometric constraints to estimate
stellar properties.

\acknowledgements

\noindent \textit{Contributions:}

The author list is divided up into 3 groups: 
\begin{enumerate}
    \item a list of primary authors who made direct 
    contributions to the construction, computation, and analysis of the
    catalog (JSS to BDJ),
    \item an alphabetized secondary list of authors who made direct or
    indirect contributions to project and/or data products (AB to EFS), and
    \item an alphabetized tertiary list of authors who provided
    useful feedback during the development process and/or on the paper itself
    (AD to IAZ).
\end{enumerate}

\noindent \textit{Personal:}

JSS would like to thank Rebecca Bleich for
her truly incredible support -- mental, physical, emotional, and spiritual -- 
during these difficult times. Without it, this paper
(along with a great many other things) would likely
never have seen the light of day.
JSS would also like to thank Jan Rybizki, Seth Gossage, and Nayantara Mudur
for insightful discussions that improved the quality of this work
and Blakesley Burkhart for providing feedback
on earlier versions of the catalog. \\

\noindent \textit{Funding:}

JSS and CZ were partially supported by the Harvard Data Science Initiative. 
HMK  acknowledges  support  from  the  DOE  CSGF  under  grant  number DE-FG02-97ER25308. 
AD received support from the National Aeronautics and 
Space Administration (NASA) under Contract No. NNG16PJ26C issued 
through the WFIRST Science Investigation Teams Program.  
CZ and DPF acknowledge support from NSF grant AST-1614941, 
“Exploring the Galaxy: 3-Dimensional Structure and Stellar Streams.”
AKS gratefully acknowledges support by a National Science 
Foundation Graduate Research Fellowship (DGE-1745303). 
YST acknowledges financial support from the Australian Research Council
through DECRA Fellowship DE220101520. \\

\noindent \textit{Data:}

The Pan-STARRS1 Surveys and the public science archive 
have been made possible through contributions by the 
Institute for Astronomy, the University of Hawaii, the Pan-STARRS Project Office,
the Max Planck Society and its participating institutes, 
the Max Planck Institute for Astronomy, Heidelberg and the Max Planck Institute 
for Extraterrestrial Physics, Garching, The Johns Hopkins University, Durham
University, the University of Edinburgh, the Queen’s University Belfast,
the Center for Astrophysics $|$ Harvard \& Smithsonian, the Las Cumbres Observatory Global
Telescope Network Incorporated, the National Central University of Taiwan, 
the Space Telescope Science Institute, the National Aeronautics and Space Administration
under Grant No. NNX08AR22G issued through the Planetary Science Division
of the NASA Science Mission Directorate, the National Science Foundation 
Grant No. AST-1238877, the University of Maryland, Eotvos Lorand University (ELTE), 
the Los Alamos National Laboratory, and the Gordon and Betty Moore
Foundation.

This publication makes use of data products from the
Two Micron All Sky Survey, which is a joint project
of the University of Massachusetts and the Infrared Processing 
and Analysis Center/California Institute of
Technology, funded by the National Aeronautics and Space Administration 
and the National Science Foundation.

This work has made use of data from the European Space Agency (ESA) 
mission Gaia (\url{https://www.cosmos.esa.int/gaia}), 
processed by the Gaia Data Processing and Analysis Consortium (DPAC;
\url{https://www.cosmos.esa.int/web/gaia/dpac/consortium}). 
Funding for the DPAC has been provided by national institutions, in
particular the institutions participating in the Gaia Multilateral Agreement.

This publication makes use of data products from the
Wide-field Infrared Survey Explorer, which is a joint
project of the University of California, Los Angeles, and
the Jet Propulsion Laboratory/California Institute of
Technology, and NEOWISE, which is a project of the Jet
Propulsion Laboratory/California Institute of Technology. 
WISE and NEOWISE are funded by the National
Aeronautics and Space Administration. \\

\noindent \textit{Computation:}

The bulk of the computation in this paper was done on the 
\textit{Cannon} cluster, which is supported by the Research Computing Group 
in the FAS Division of Science at Harvard University. \\

\noindent \textit{Code}:

This work has benefited from the following packages:
\begin{itemize}
    \item {\astropy} \citep{astropy+13,astropy+18}
    \item {\numpy} \citep{vanderwalt+11}
    \item {\scipy} \citep{virtanen+20}
    \item {\matplotlib} \citep{hunter07}
    \item {\healpy} \citep{gorski+05,zonca+19}
    \item {\gala} \citep{price-whelan17}
    \item {\galpy} \citep{bovy15}
    \item {\corner} \citep{foremanmackey16}
\end{itemize}

\bibliography{augustus}

\appendix

\section{Data Products} \label{ap:catalog}

The output {\augustus} stellar parameter catalogs 
can be found online through the 
\href{https://dataverse.harvard.edu/dataverse/brutus_augustus}{Harvard Dataverse}. 
Two types of data products are made available:
\begin{enumerate}
    \item a ``point'' catalog that contains information
    about each object along with sumamry statistics
    describing the results (\href{https://doi.org/10.7910/DVN/WYMSXV}{doi:10.7910/DVN/WYMSXV}) 
    and
    \item a ``samples'' catalog that contains 25 random
    posterior samples for each object 
    (\href{https://doi.org/10.7910/DVN/530UYQ}{doi:10.7910/DVN/530UYQ}).
\end{enumerate}
A summary of the column names, the data format,
and a brief description of each catalog can be found
in Tables \ref{tab:point_cat} and \ref{tab:samples_cat}.
Note that the ``samples'' catalogs is strictly
meant to be supplementary to the ``point'' catalog 
and is matched to the latter row-wise.\footnote{As described in \S\ref{sec:results},
\textsc{brutus} \texttt{v0.7.5} contained a bug 
(fixed in more recent versions of the code)
that used the wrong sign when sampling
from correlations between $A_V$ and $R_V$ with distance $d$. While
this has a negligible impact on the quality of the point catalog outside
of the provided random draw,
it does affect quantities computed directly
from the ``samples'' catalog which jointly depend on $d$ and $A_V$ or $R_V$
(e.g., reddened photometry).}
Catalogs are made available for sources modeled using all bands
as well as excluding UKIDSS data for sources that
have them (which have the \texttt{\_noukidss} suffix). 
The \texttt{\_noukidss} data products are available for download 
at the same \href{https://dataverse.harvard.edu/dataverse/brutus_augustus}{Harvard Dataverse} 
repository as their UKIDSS-included counterpart files.
Additional information on the columns provided in these catalogs
are described below.

Information on sources is provided through their corresponding Pan-STARRS ID
(\texttt{PS\_ID}) and \textit{Gaia} DR2 ID (\texttt{GAIA\_ID}) as well as by
their 2-D coordinates, in units of right ascension and declination
(\texttt{SKY\_COORDS}) as well as Galactic longitude and latitude
(\texttt{GAL\_COORDS}). In addition, we also include
astrometric measurements from \textit{Gaia} including parallaxes
(\texttt{PARALLAX}, \texttt{PARALLAX\_ERROR}) and proper motions
(\texttt{PROPER\_MOTION}, \texttt{PROPER\_MOTION\_ERROR)} as well as multi-band
photometry from \textit{Gaia}, Pan-STARRS, 2MASS, UKIDSS, and unWISE
(\texttt{MAGNITUDES}, \texttt{MAGNITUDES\_ERROR}). Note that the
parallaxes and magnitudes contain \textit{none} of the offsets or additional
systematic corrections/errors described in \S\ref{sec:data} outside of the $0.005\,{\rm mag}$
photometric error floor.

Information on the overall quality of the fit can be assessed in a few ways.
One metric is the log-evidence (\texttt{LOG\_EVID}), defined as
\begin{equation}
    \ln \mathcal{Z} \equiv \ln\left( \int
    \likelihood_{\rm phot}(\params, \eparams) \,
    \likelihood_{\rm astr}(\eparams) \,
    \prior(\params, \eparams) \, 
    \deriv\params \, \deriv\eparams \right)
\end{equation}
We estimate this for each object by summing over 
the final (weighted) subset of models before applying
the posterior resampling scheme described in Speagle et al. (2021a, subm.).
This provides information on the \textit{overall} quality of the fit
across all the models \textit{including} the influence of the prior.

Another metric is simply the best-fit $\chi^2_{\rm best}$ (\texttt{BEST\_CHI2}) 
from all the models before applying the posterior resampling scheme.
This provides information on the quality of the \textit{best} possible
fit \textit{ignoring} the impact of the prior, 
thereby serving as a useful supplement to the log-evidence. 

Combined with the number of bands $b$ used in the fit (\texttt{NBANDS\_IN\_FIT}),
we use this information to derive a flag for rejecting sources
that fail to achieve even a single reasonable fit:
\begin{equation}
\texttt{FLAG\_FIT} =
    \begin{cases}
    \texttt{TRUE} & {\rm if}\:P(\chi^2>\chi^2_{\rm best}|b-3) < 10^{-3} \\
    \texttt{FALSE} & {\rm otherwise}
    \end{cases}
\end{equation}
where $P(\chi^2>\chi^2_{\rm best}|b-3)$ is the probability of observing
a $\chi^2$ value larger than $\chi^2_{\rm best}$ assuming $b-3$ degrees of
freedom. Note that we use $b-3$ rather than $b$ due to the fact that
{\brutus} ``optimizes'' over 3 parameters ($d$, $A_V$, $R_V$) before
the posterior weighting and resampling step.

For every parameter we compute the 2.5th, 50th, and 97.5th percentiles
(i.e. the median and the $2\sigma$ errors) for each parameter
by rank-ordering the final set of $n=250$ posterior samples. We choose to report $2\sigma$ rather than
$1\sigma$ errors here since they better reflect intuitive understanding of uncertainty
(i.e. objects are ``unlikely'' to be outside the errors bars) and better highlight
possible degeneracies in the fits (e.g., between dwarf and giant solutions) when they occur.
We provide these percentiles along with a random sample taken
from the posterior for the object for each parameter in our model 
(see Table \ref{tab:point_cat} for a full list). Since all the parameters from this
random sample are correlated, it can be useful in certain contexts.

We use these percentiles to define a second flag \texttt{FLAG\_GRID}
that we set to \texttt{TRUE} if any of the 2.5th or 97.5th percentiles for each parameter 
that defines the grid of models ($M_{\rm init}$, $\feh_{\rm init}$, ${\rm EEP}$) are equal
to the minimum or maximum possible value of that parameter, respectively, and
\texttt{FALSE} otherwise. This flags posteriors that may be biased due to the
hard edges present in our input model grid. This mainly flags
sources with lower initial masses since our model grid only goes down to
$M_{\rm init} = 0.5\,M_\odot$.

Finally, as part of the ``point'' catalog we also provide 
the probability \texttt{PROB\_GIANT} that a source
is a giant, which we define as
\begin{equation}
    \texttt{PROB\_GIANT} \equiv P(\log g < 3.5)
\end{equation}
following the definition used in the H3 survey \citep{conroy+19a}.
We estimate this using the final set of posterior samples, which gives us
a resolution of $1/n_{\rm samp} = 1/250 = 0.4\%$ in probability.

In the ``samples'' catalog, we provide 25 samples from the posterior
for the distance $d$ (\texttt{SAMPLES\_DISTANCE}), extinction $A_V$
(\texttt{SAMPLES\_A\_V}), differential reddening $R_V$
(\texttt{SAMPLES\_R\_V}), and ``model index''
(\texttt{SAMPLES\_MODEL\_IDX}). The model index can be used to grab the
corresponding models from the input parameter grid, which can then
be used to construct output predictions for associated quantities.

An example showing how to use these samples within {\brutus} is shown below:

\begin{verbatim}
import numpy as np
import h5py
from brutus import utils as butils
from brutus.filters import gaia, ps, tmass, ukidss, wise

# grab quality flags
cat = h5py.File(`point_cat.h5', mode=`r')  # load h5 file
flag_fit, flag_grid = cat[`FLAG_FIT'][:100], cat[`FLAG_GRID'][:100]  # first 100 elements
good = np.where(~flag_fit & ~flag_grid)[0]  # no flags (good fits, good posteriors)

# load samples catalog
samples = h5py.File(`samples_cat.h5', mode=`r')  # load h5 file
samples_idx = samples[`SAMPLES_MODEL_IDX'][:100][good]  # first 100 elements + no flags

# load MIST grid
flts = gaia + ps[:-2] + tmass + ukidss + wise[:-2]  # define filterset
mags, labels, _ = butils.load_models(`grid_mist_v8.h5', filters=flts)  # read file

# get effective temperatures of corresponding models
logt = labels[`logt'][samples_idx]  # log(Teff)
teff = 10**logt  # convert from log to linear

# compute percentiles (median, +/- 1 sigma, +/- 2 sigma)
teff_vals = np.percentile(teff, [2.5, 16, 50, 84, 97.5], axis=1)

# compute mean and standard deviation of predicted intrinsic Gaia G magnitude at 1 kpc
G = mags[:, 0, 0][samples_idx]
G_mean, G_std = np.mean(G, axis=1), np.std(G, axis=1)
\end{verbatim}

\begin{deluxetable*}{lll}
\tablecolumns{3}
\tablecaption{Summary of the {\augustus} ``point'' catalog that includes
object information and summary statistics description results
from the {\brutus} fits. See Appendix \ref{ap:catalog} for additional
details. The table is available for download at 
\href{https://doi.org/10.7910/DVN/WYMSXV}{doi:10.7910/DVN/WYMSXV}.
\label{tab:point_cat}}
\tablehead{Name & Data Format & Description}
\startdata
\cutinhead{\textbf{Object Information}}
\texttt{PS\_ID} & 64-bit uint & Pan-STARRS object ID \\
\texttt{GAIA\_ID} & 64-bit uint & \textit{Gaia} DR2 object ID \\
\texttt{SKY\_COORDS} & 64-bit float (x2) & Sky coordinates $(\alpha, \delta)$ in degrees \\
\texttt{GAL\_COORDS} & 64-bit float (x2) & Galactic coordinates $(\ell, b)$ in degrees \\
\texttt{PARALLAX} & 32-bit float & Parallax from \textit{Gaia} DR2 in mas \\
\texttt{PARALLAX\_ERROR} & 32-bit float & Parallax error from \textit{Gaia} DR2 in mas \\
\texttt{PROPER\_MOTION} & 32-bit float (x2) & Proper motion in sky coordinates from \textit{Gaia} DR2 in mas/yr \\
\texttt{PROPER\_MOTION\_ERROR} & 32-bit float (x2) & Proper motion error in sky coordinates from \textit{Gaia} DR2 in mas/yr \\
\texttt{MAGNITUDES} & 32-bit float (x16) & Magnitudes from \textit{Gaia} DR2, Pan-STARRS, 2MASS, UKIDSS, and unWISE \\
\texttt{MAGNITUDES\_ERROR} & 32-bit float (x16) & Magnitude errors from \textit{Gaia} DR2, Pan-STARRS, 2MASS, UKIDSS, and unWISE \\
\cutinhead{\textbf{Fit Information}}
\texttt{LOG\_EVID} & 32-bit float & Log-evidence (base $e$) from models used in the fit \\
\texttt{BEST\_CHI2} & 32-bit float & Best-fit $\chi^2$ value (photometry and parallax) from models used in the fit \\
\texttt{NBANDS\_IN\_FIT} & 8-bit uint & Number of bands (photometry and parallax) included in the fit \\
\texttt{FLAG\_FIT} & 1-bit bool & Whether there was an issue with the fit (\texttt{TRUE} = yes) \\
\texttt{FLAG\_GRID} & 1-bit bool & Whether the posterior hits the edge of the grid (\texttt{TRUE} = yes) \\
\cutinhead{\textbf{Stellar Properties}}
\texttt{PROB\_GIANT} & 16-bit float & Probability that $\log g < 3.5$ from models used in the fit \\
\texttt{INIT\_MASS} & 16-bit float (x4) & 2.5th, 50th, and 97.5th percentiles and a random sample of $M_{\rm init}$ in $M_\odot$ \\
\texttt{INIT\_FEH} & 16-bit float (x4) & 2.5th, 50th, and 97.5th percentiles and a random sample of $\feh_{\rm init}$ \\
\texttt{EEP} & 16-bit int (x4) & 2.5th, 50th, and 97.5th percentiles and a random sample of ${\rm EEP}$ \\
\texttt{LOG10\_AGE} & 16-bit float (x4) & 2.5th, 50th, and 97.5th percentiles and a random sample of $\log t_{\rm age}$ in yr \\
\texttt{LOG10\_TEMP\_EFF} & 16-bit float (x4) & 2.5th, 50th, and 97.5th percentiles and a random sample of $\log T_{\rm eff}$ in K \\
\texttt{LOG10\_LBOL} & 16-bit float (x4) & 2.5th, 50th, and 97.5th percentiles and a random sample of $\log L_{\rm bol}$ in $L_\odot$ \\
\texttt{LOG10\_SURF\_GRAV} & 16-bit float (x4) & 2.5th, 50th, and 97.5th percentiles and a random sample of $\log g$ in cgs \\
\texttt{DISTANCE} & 32-bit float (x4) & 2.5th, 50th, and 97.5th percentiles and a random sample of $d$ in kpc \\
\texttt{A\_V} & 16-bit float (x4) & 2.5th, 50th, and 97.5th percentiles and a random sample of $A_V$ in mag \\
\texttt{R\_V} & 16-bit float (x4) & 2.5th, 50th, and 97.5th percentiles and a random sample of $R_V$ \\
\enddata
\end{deluxetable*}

\begin{deluxetable*}{lll}
\tablecolumns{3}
\tablecaption{Summary of the {\augustus} ``samples'' catalog that includes
random posterior samples from the {\brutus} fits.
See Appendix \ref{ap:catalog} for additional details.
The table is available for download at 
\href{https://doi.org/10.7910/DVN/530UYQ}{doi:10.7910/DVN/530UYQ}.
\label{tab:samples_cat}}
\tablehead{Name & Data Format & Description}
\startdata
\texttt{SAMPLES\_MODEL\_IDX} & 32-bit int (x25) & 25 posterior samples of the model index in the input grid \\
\texttt{SAMPLES\_DISTANCE} & 32-bit float (x25) & 25 posterior samples of $d$ in kpc \\
\texttt{SAMPLES\_A\_V} & 16-bit float (x25) & 25 posterior samples of $A_V$ in mag \\
\texttt{SAMPLES\_R\_V} & 16-bit float (x25) & 25 posterior samples of $R_V$ in mag \\
\enddata
\end{deluxetable*}

\end{document}